\documentclass[12pt]{article}

\input{epsf}
\usepackage[dvips]{graphicx}
\usepackage{cite}
\usepackage{color}
\def\1ad{\mbox{\normalsize $^1$}}
\def\2ad{\mbox{\normalsize $^2$}}
\def\3ad{\mbox{\normalsize $^3$}}
\def\4ad{\mbox{\normalsize $^4$}}

\def\5ad{\mbox{\normalsize $^5$}}
\def\6ad{\mbox{\normalsize $^6$}}
\def\7ad{\mbox{\normalsize $^7$}}
\def\8ad{\mbox{\normalsize $^8$}}

\def\dj{\hbox{d\kern-0.347em \vrule width 0.3em height 1.252ex depth
-1.21ex \kern 0.051em}}

\newcommand{\be}{\begin{equation}}
\newcommand{\ee}{\end{equation}}
\newcommand{\ben}{\begin{equation*}}
\newcommand{\een}{\end{equation*}}
\newcommand{\ba}{\begin{eqnarray}}
\newcommand{\ea}{\end{eqnarray}}
\newcommand{\ban}{\begin{eqnarray*}}
\newcommand{\ean}{\end{eqnarray*}}
\newcommand{\brr}{\begin{array}}
\newcommand{\err}{\end{array}}
\newcommand{\bc}{\begin{center}}
\newcommand{\ec}{\end{center}}

\newcommand{\bea}{\begin{eqnarray}}
\newcommand{\eea}{\end{eqnarray}}
\newcommand{\bean}{\begin{eqnarray*}}
\newcommand{\eean}{\end{eqnarray*}}

\newcommand\lsim{\mathrel{\rlap{\lower4pt\hbox{\hskip1pt$\sim$}}
    \raise1pt\hbox{$<$}}}
\newcommand\gsim{\mathrel{\rlap{\lower4pt\hbox{\hskip1pt$\sim$}}
    \raise1pt\hbox{$>$}}}

\begin{document}

\setcounter{page}{0}
\thispagestyle{empty}

\begin{flushright}
CERN-PH-TH/2006-133\\
hep-ph/0607158
\end{flushright}

\vskip 8pt

\begin{center}
{\bf \Large {

Gravitational Waves\\ [0.25cm]
from Warped Spacetime
}}
\end{center}

\vskip 10pt

\begin{center}
{\large Lisa Randall$^{a}$ and G\'eraldine Servant $^{b,c}$ }
\end{center}

\vskip 20pt

\begin{center}

\centerline{$^{a}$ {\it Jefferson Physical Laboratory, Harvard University, Cambridge, MA, 02138}}
\vskip 3pt
\centerline{$^{b}${\it Service de Physique Th\'eorique, CEA Saclay, F91191 Gif--sur--Yvette,
France}}
\vskip 3pt
\centerline{$^{c}${\it CERN Physics Department, Theory Division, CH-1211 Geneva 23, Switzerland}}
\vskip .3cm
\centerline{\tt  randall@physics.harvard.edu, geraldine.servant@cern.ch}
\end{center}

\vskip 13pt

\begin{abstract}
\vskip 3pt
\noindent

We argue that the RSI model can provide a strong signature in gravitational waves. This signal is a relic stochastic background generated during the cosmological phase transition from an AdS-Schwarschild phase to the RS1 geometry that should occur at a temperature in the TeV range. We estimate the amplitude of the signal in terms of the parameters of the potential stabilizing the radion and show that over much of the parameter region in which the phase transition completes, a signal should be detectable at the planned space interferometer, LISA.

\end{abstract}

\vskip 13pt
\newpage

\section{Introduction}

The nature of the theory underlying the electroweak phase transition  will
hopefully be resolved within the next five
 years at the LHC.   However, other indirect probes of the weak scale could
supplement these results and provide
further important insights. The nature of the dark matter, for example, might
provide weak scale information. In this light, the planned LISA gravitational wave detector
could be very exciting since the frequencies of the observable gravitational waves lie in just the right range for
exploring the electroweak scale. In this letter, we apply this observation to the
warped five-dimensional spacetime of
the RS1 model and show that the early universe phase transition between the
AdS-Schwarshild and RS1 phases could
 provide a sizable signal.

The frequency of gravitational waves observed today is
\be
f=f_* \frac{a_*}{a_0}=f_* \left( \frac{g_{s0}}{g_{s*}}\right)^{1/3}\frac{T_0}{T_*}  \approx 6 \times 10^{-3} \mbox{mHz} \left(\frac{g_*}{100}\right)^{1/6}\frac{T_*}{100\mbox{ GeV}} \frac{f_*}{H_*}
\ee
where $f_*$, $T_*$, $H_*$, $g_*$ are respectively the characteristic
frequency, temperature, Hubble frequency and
 number of relativistic degrees of freedom at the time when the gravitational waves were produced.
For weak scale temperatures and
$f_*/H_*\sim 10^2$ (as expected for weak scale phenomena as we explain
below), this is peaked in the
LISA band ($10^{-4}-10^{-2}$) Hz and is actually a stronger signal for the
range slightly outside that probed
 by the LHC. That is, even if weak scale physics is at the high end of the
LHC range, we might probe the
 underlying theory further at LISA.

However, not all weak scale physics is relevant to LISA observations since
only rather dramatic dynamical
 phenomena will give rise to detectable gravitational waves. One such
possibility is a sufficiently strong
 first order phase transition. In this letter, we point out that  if the
RS1 warped geometry resolution to the
 hierarchy problem \cite{Randall:1999ee} is correct, it could yield strong
detectable signals.
This would be significant not only as a potential confirmation of LHC results, but is
also very promising because
 the potential energy reach of the gravitational wave signal is higher, and
phase transitions up to 10 TeV
 might yield visible signals, while the second generation space
interferometer Big Bang Observer (BBO)
 can explore even higher energies up to $10^7$ GeV \cite{Grojean:2006bp}. Furthermore, the LISA detector would probe the early universe phases, potentially supplementing whatever might be learned at the LHC.

The key point is that at high temperature, there is an AdS-Schwarschild phase involving a single brane where the graviton amplitude is peaked, whereas at low energies, there are two branes with a slice of bulk AdS in between \cite{Randall:1999ee}. In Ref. \cite{Creminelli:2001th}, it was shown that in the perturbative regime one expects a first order phase transition between these two phases, which  proceeds through the nucleation of ``brane bubbles''.  Ref. \cite{Creminelli:2001th} focused on the difficulty of completing this phase transition in the perturbative regime consistently with small back-reaction, finding their analysis favored lower $N$ (where $N$ is the parameter of the conformal field theory that determines the ratio of the five-dimensional Planck scale and the  AdS scale).  In this letter, we discuss some aspects of these results, considering also the negative $\epsilon$ case (essentially the squared mass of the Goldberger-Wise scalar living in the AdS bulk), and demonstrate the  potential detectability of the gravitational wave signal associated with the first order phase transition over a  large region of parameter space.

Our analysis follows closely the methodology of
 Refs. \cite{Kosowsky:1992rz,Kosowsky:1991ua,Kosowsky:1992vn
,Kamionkowski:1993fg,Kosowsky:2001xp,Dolgov:2002ra,Apreda:2001us,Nicolis:2003tg}
 which applies to
very strong phase transitions (like the phase transition we are considering
 in this work, as
 will be shown shortly), in which case bubble expansion proceeds via
 detonation \cite{Kamionkowski:1993fg}.
In this regime, the gravitational wave signal depends only on two
 parameters. The first is the dimensionless parameter $\alpha$, which is
 defined as the ratio of the
 latent heat to the radiation energy density evaluated at the nucleation
 temperature $T_n$.
To achieve a visible signal at LISA, $\alpha$ must be sufficiently big, at least 0.2 \cite{Nicolis:2003tg,Grojean:2006bp}.

The second parameter, $\beta$, tells the time variation of the bubble
 nucleation rate and hence the length of
 time of the phase transition. It is defined as $\beta/H=T d(S_3/T)/dT$
 where $S_3$ is the free energy of the
 bubble and the derivative is also evaluated at the nucleation
 temperature. A large signal requires a relatively slow
phase transition so $\beta/H$ should be small. A visible signal at LISA
 requires $\beta/H\lsim10^3$. $\beta/H$
 is dimensionless and its size is mainly determined by the shape of the
 effective potential
 at the nucleation temperature $T_n$. It depends on the energy scale $T_n$
 only logarithmically.
Typically, $\beta / H \sim S_3/T$. Nucleation occurs when the probability
 to nucleate one bubble per
 horizon time and horizon volume  $\sim T^4 e^{-S_3/T}/H^4$ becomes of
 order unity.
 Therefore the value of $S_3/T$ at $T_n$ is about $ 4 \ln
 (m_{Pl}/T_n)$. For $T_n\sim 10^2-10^3$ GeV,
 we then expect $\beta/H \sim {\cal O}(10^2)$ which is in the visible range.

There are two sources of gravitational waves from a first-order phase
 transition: bubble collisions and turbulence
 in the plasma.  The corresponding relic signals, expressed in terms of the
 fraction of the total energy
 density today, at the peak frequencies $f_{\mbox{\tiny coll}}$ and $f_{\mbox{\tiny turb}}$, are
 \cite{Kamionkowski:1993fg,Kosowsky:2001xp,Dolgov:2002ra, Nicolis:2003tg} :
\bea
\label{Omegacoll}
\Omega_{\mbox{\tiny coll}} \ h_0^2 (f_{\mbox{\tiny coll}}) & \simeq &1.1 \times 10^{-6} \kappa^2 \left[\frac{H_*}{\beta}\right]^2 \left[\frac{\alpha}{1+\alpha}\right]^2\left[\frac{v_b^3}{0.24+v_b^3}\right]\left[\frac{100}{g_*}\right]^{1/3}\\
\label{fcoll}
f_{\mbox{\tiny coll}} &\simeq &5.2 \times 10^{-3} \mbox{mHz} \left[\frac{\beta}{H_*}\right]\left[\frac{T_*}{100 \mbox{GeV}}\right]\left[\frac{g_*}{100}\right]^{1/6}\\
\label{Omegaturb}
\Omega_{\mbox{\tiny turb}} \ h_0^2 (f_{\mbox{\tiny turb}})& \simeq &1.4 \times 10^{-4} u_s^5v_b^2 \left[\frac{H_*}{\beta}\right]^2 \left[\frac{100}{g_*}\right]^{1/3}\\
\label{fturb}
f_{\mbox{\tiny turb}} &\simeq &3.4 \times 10^{-3} \mbox{mHz} \frac{u_s}{v_b}\left[\frac{\beta}{H_*}\right]\left[\frac{T_*}{100 \mbox{GeV}}\right]\left[\frac{g_*}{100}\right]^{1/6}
\eea
 $u_s$ and $\kappa$ are respectively the bubble wall velocity, the turbulent fluid velocity and the fraction of vacuum energy which goes into kinetic energy of bulk motions of the fluid. They are given by \cite{Steinhardt:1981ct,Kamionkowski:1993fg,Nicolis:2003tg}
\bea
\nonumber
v_b(\alpha)=\frac{1/\sqrt{3} +\sqrt{\alpha^2+2\alpha/3}}{1+\alpha}   \   , \ \ u_s(\alpha)\simeq \sqrt{\frac{\kappa \alpha}{\frac{4}{3}+\kappa \alpha}} \  , \
\kappa(\alpha)\simeq \frac{1}{1+0.715 \alpha}\left[0.715 \alpha +\frac{4}{27}\sqrt{\frac{3\alpha}{2}}\right]
\eea
Formulae (\ref{Omegaturb}), (\ref{fturb}) for turbulence were recently corrected in \cite{Caprini:2006jb} and they are being revisited for the collision signal \cite{CDS}. However, for the purpose of the present paper, we use Eqs. (\ref{Omegacoll}), (\ref{fcoll}),
(\ref{Omegaturb}), (\ref{fturb}).

The phase transition occurs when the radion that determines the distance between the two branes of RS1 is stabilized.
Note that the original description of RS1 is five-dimensional, yet we are using a four-dimensional formalism to determine the strength of the phase transition and the gravitational wave signal. This is justified in the regime where the radion is light (lighter than the KK modes) so that it dominates the potential in the RS regime, which requires that the parameters determining the radion potential are perturbative. In the high temperature phase, the AdS-Schwarschild metric can be interpreted holographically in terms of a four-dimensional CFT, so one can use the four-dimensional formalism in that regime as well.

To evaluate the relevant quantities, we use the formalism of Creminelli et al \cite{Creminelli:2001th}, who
assumed the original Goldberger-Wise (GW) model in which there is a
scalar field whose bulk mass $m$ is
 determined by the parameter $\epsilon=\sqrt{4+m^2/k^2} -2$. This field
takes values $v_0$ on the Planck brane
and $v_1$ on the TeV brane. From the five-dimensional perspective, the
radion is stabilized by the tension
between the mass term and the gradient contribution to the
energy. Because
 of the background anti de Sitter space, $\epsilon \approx m^2/4k^2$  can be positive or
negative, though the original paper
 considered only the positive case\cite{Goldberger:1999uk}. Other variations on this model include
interactions of the scalar field but we don't expect this to change the results significantly. 

  One can also consider the holographic
interpretation of this model
 \cite{Arkani-Hamed:2000ds,Rattazzi:2000hs}. In this interpretation one has a marginal operator, which is relevant for positive $\epsilon$ and irrelevant for negative $\epsilon$. The positive case is peculiar from a holographic perspective. In this interpretation  the breaking of conformal symmetry in the IR occurs because of the competing effect of two terms. The negative $\epsilon$ case is more general and conventional, and corresponds to a coupling that grows in the IR, where  it breaks the conformal symmetry, analogously to the  QCD phase transition. We will consider both positive and negative $\epsilon$.

The five-dimensional metric is $ds^2=e^{-2ky} \eta_{\mu \nu}dx^{\mu}
dx^{\nu}+dy^2$ where $k=1/L$ is the
 AdS curvature. The kinetic term for the radion field  $\mu=k e^{-ky}$, in
terms of the five-dimensional
 Planck mass $M$, is \cite{Rattazzi:2000hs,Csaki:1999mp,Goldberger:1999un}
\be
{\cal L}_{kin}=-12\sqrt{-g}(ML)^3(\partial \mu)^2
\label{kinetic}
\ee
and its induced four-dimensional potential is
\cite{Goldberger:1999uk}
\be
V_{\rm GW}(\mu) = \epsilon v_0 ^2 \mu_0 ^4 +
\left[(4+2\epsilon)\mu^4(v_1-v_0(\mu/\mu_0)^\epsilon)^2-\epsilon v_1^2\mu^4 +\delta T_1 \mu^4\right]
+ {\cal{O}} (\mu^8 / \mu_0 ^4)          \; ,
\end{equation}
where $\mu_0 $ is the UV scale and  $| \epsilon | \ll 1$ has been
assumed. The terms $v_0$ and $v_1$ are the vevs in Planck units of the Goldberger-Wise field on the Planck and TeV branes.

We sometimes modify the potential of \cite{Goldberger:1999uk} to allow a
 term $\delta T_1 \mu^4$ corresponding to
 a change of the TeV brane tension.
Such a term is permissible and allows for a larger range of viable models.

Provided that  $ \delta T_1 < \epsilon v_1^2$ the potential above has a global minimum at
\bea
\label{mutev}
\mu_{\mbox{\tiny TeV}\pm}
&\approx& \mu_0\left(\frac{v_1}{v_0}\right)^{1/\epsilon}\left[1\pm \frac{1}{2}\sqrt{-\frac{\delta T_1}{v_1^2}+\epsilon} \right]^{1/\epsilon}
\eea
where $\pm$ corresponds to the cases $\epsilon >0$ and $\epsilon < 0$ respectively.
 For $\epsilon <0$, we have the additional condition $ \delta T_1>-v_1^2(4+\epsilon)$ so that the minimum above does not become a maximum.
 For $\epsilon > 0$ and  $\delta T_1 = 0$, $ \mu_{\mbox{\tiny TeV}\pm}\sim
\mu_0 \left({v_1}/{ v_0}\right)^{{1}/{\epsilon}}$. In that case,
the hierarchy between the weak and Planck scale can be naturally obtained
for parameters not far from 1.
For $\epsilon<0$ and $\delta T_1 = 0$ the only minimum of the potential is
at $\mu=0$. With nonzero $\delta T_1 <0 $  there is also the desired
minimum at $\mu \sim \mu_0 (v_0/v_1)^{1/|\epsilon|}  (1-\sqrt{\delta T_1}/2v_1)^{1/|\epsilon|}$ and a viable solution can be obtained for $v_0<v_1$ \cite{Arkani-Hamed:2000ds,Rattazzi:2000hs}.
 Note that $\mu=0$ is also a minimum of the potential but for $\epsilon>0$
 and  $ \delta T_1<-v_1^2(4+\epsilon)$,  the barrier disappears and $\mu=0$ becomes a maximum.

For small $\epsilon$, the value of the potential at the
minimum (\ref{mutev}) where the radion achieves its vacuum expectation value is
\be
V_{\pm}\approx   \mu_{\mbox{\tiny TeV}\pm}^4 \ {\epsilon}( \delta T_1/2 \mp v_1 \sqrt{-\delta T_1 + \epsilon v_1^2})
\ee
where we have dropped $\epsilon v_0^2 \mu_0^4$ which is common to the high
and low energy phases and won't contribute to $\alpha$ or $\beta$. \\

When $\delta T_1=0$, $\epsilon$ has to be positive.
 The value of the potential at the minimum scales like $- \epsilon^{3/2} v_1^2 \mu_{\mbox{\tiny TeV}}^4$,
where $v_1$ is the assumed value of the Goldberger-Wise field on the TeV brane.
As the free energy of a critical bubble is smaller if the minimum is deep, it will clearly be smaller when $v_1$ is
larger. However, overly large $v_1$ could result in a large back-reaction to the potential as we will see in Section \ref{sec:perturbativity}.  A quick way to extrapolate the $\delta T_1=0$ results for large $\delta T_1$ is to note that the potential at the minimum is then proportional to $\mu_{\mbox{\tiny TeV}}^4 \epsilon (\delta T_1/2 \mp v_1 \sqrt{- \delta T_1})$. In other words, the quantity $\epsilon^{3/2} v_1^2$ which appeared in the minimum energy for $\delta T_1=0$ is replaced by $\epsilon \delta T_1 v_1^2$. Clearly, larger $|\delta T_1|$ yields a deeper minimum. This makes the tunneling amplitude bigger.

\section{Completion of the phase transition}

The free energy of the AdS-S phase is \cite{Creminelli:2001th}
\be
F_{\mbox{\tiny AdS-S}}=-2 \pi^4 (ML)^3T^4
\label{FAdSS}
\ee
  (\ref{FAdSS})  is a minimum of the free energy of AdS-S corresponding to the Hawking temperature at the horizon of the black hole solution equal to the   temperature of the universe. By
holography, $F_{\mbox{\tiny AdS-S}}$ can be interpreted as the free energy
of a strongly coupled large $N$ CFT with $N^2=16\pi^2(ML)^3$. The exact relation depends on the precise theory; this formula is for ${\cal N}=4$ $SU(N)$ super Yang Mills \cite{Gubser:1999vj}.  We often choose to use the
parameter $N$ to characterize the AdS curvature.

In the Randall-Sundrum phase at high temperature, the TeV brane is pushed by thermal effects to the AdS horizon in the absence of the Goldberger-Wise field. Since it costs energy for $\mu$ to be nonzero, there exists an energy barrier which leads to a first-order phase transition.
The phase transition can take place after the energy of the AdS-S phase equals that of the RS phase.
Another argument for the first order phase transition is provided by the AdS-CFT correspondence that relates the RS model with a 4D confining gauge theory. It is well-known that the confining phase transition of large $N$ ($N\gsim 3$) gauge theories is first order (growing more strongly as $N$ is large). In the 4D description, the stabilized radion is some glueball state.
The critical temperature is the temperature at which the RS energy at the minimum of the GW potential equals the AdS-Schwarschild solution energy. We have
\be
T_c=\left(\frac{-8V_{\pm}}{\pi^2N^2}\right)^{1/4}
\ee
The phase transition can proceed only if the  bubble nucleation rate is
larger than the expansion rate of the universe i.e.  $ S_3/T \lsim 4 \ln
(m_{Pl}/T_n)$ where $S_3$ is the free energy of a critical bubble. For
electroweak
 scale temperatures, this corresponds to the condition $ S_3/T \lsim 140$.
The latent heat is of the order of $V_{\pm}$.

We now follow the assumption of  \cite{Creminelli:2001th}, where
the contribution from the AdS-S side to the thermal bounce action is neglected. This is reasonable when $v_1$ and
$\epsilon$ are small and the
 potential is consequently shallow, in which case the bubbles are big and most
of the action comes from
 the RS regime as $\mu$ changes from 0 to $\mu_{TeV}$. We therefore integrate
only over the RS side
 when evaluating the action.  This approximation is reasonable
provided that
 $T_c \ll \mu_{\mbox{\tiny TeV}\pm}$ in which case only the radion mode
contributes to the action and is valid so long as $v_1$ is small\footnote{Before the holographic description was emphasized, a study of the phase transition with a different radion potential (and  a different initial thermal phase) was performed in \cite{Cline:2000xn}.}.

The quantity $\alpha$ depends on the  free energy difference between the two minima, which is
\be
 \Delta V_{\pm}(\mu)=V_{GW}(\mu) + \frac{\pi^2N^2T^4}{8}=
V_{\pm}\left[\frac{V_{GW}(\mu)}{V_{\pm}}-\left(\frac{T}{T_c}\right)^4\right]
\ee
where we have deliberately left open the possibility that at nucleation $V_{GW}(\mu)$ is not exactly the value of the potential at the minimum $V_{\pm}$.

The quantity $\beta$ is determined from  the bubble action, which
 is greatly simplified in two limiting cases: the thin-wall and thick-wall
approximations. In the first case,
the bubble radius is much larger than the thickness of the wall (region
over which the value of $\mu$ varies). This applies when
$\Delta V_{\pm}$ is much smaller
 than the height of the barrier separating the two minima (which we do not know).

However, if the thin-wall action is too big to allow nucleation, the action proceeds via
thick-wall bubbles.
 As the temperature drops, $\Delta V_{\pm}$ increases and eventually
becomes larger than the barrier.
 It is then favorable to make the wall thickness comparable to the bubble
size to minimize the surface term.
We will soon see that the phase transition occurs primarily in the thick-wall regime in which supercooling is relatively large.

The tunneling rate is proportional to ${\rm exp}(-S)$, where $S=S_3/T$ if thermal
bubbles dominate, where $S_3=4\pi\int r^2dr[(\partial\mu/\partial
r)^2/2+\Delta V(\mu)]$ is the 3D Euclidean action of an $O(3)$-symmetric critical bubble, and
$S=S_4=2\pi^2 \int r^3dr[(\partial\mu/\partial r)^2/2+\Delta V(\mu)]$ in the regime
where $O(4)$-symmetric bubbles dominate.

In general, nucleation proceeds via thermal bubbles but if the phase transition doesn't complete before a temperature $T$ such that $T\lsim
(2R)^{-1}$, where $R$ is the radius of the $O(4)$
bubble, and $S_4<S_3/T$
the $O(4)$ symmetric solution \cite{Linde:1981zj} applies. We find that $S_4$
is never smaller than
$S_3/T$ in the thin wall approximation. However, for thick wall, there will
be some region of
parameter space where $S_4$ is more favorable.

In the thin-wall approximation \cite{Linde:1981zj},
\be
S_3= ({16 \pi}/{3}) \left({3 N^2}/{2\pi^2}\right)^{3/2} {S_1^3}/{( \Delta V_{\pm}(\mu))^2}
\label{S3thin}
\ee
where  $S_1=\int_0^{ \mu_{\mbox{\tiny TeV} \pm}} d\mu \sqrt{2|\Delta V|}\approx -\int_0^{ \mu_{\mbox{\tiny TeV} \pm}} d\mu \sqrt{-2V_{\pm}}$ is the surface tension evaluated in the limit $T\rightarrow T_c$ and the  $({3N^2}/{2\pi^2})^{3/2}$ factor comes from the canonical normalization of $\mu$.
We evaluate the denominator in (\ref{S3thin}) at the minimum of the potential, $\mu=\mu_{\mbox{\tiny TeV}} $, for which:
\be
 \Delta V_{\pm}=V_{\pm}(1-(T/T_c)^4)
\ee
 This leads to\footnote{This is a factor $8\times 2^{3/2}$ larger than in \cite{Creminelli:2001th}. }  $ S_3\sim({16 \pi}/{3}) 2^{3/2}({\mu^3_{\mbox{\tiny TeV}}}/{ \sqrt{| V_{\pm}|}})\left({3N^2}/{2\pi^2}\right)^{3/2} /{(1-(T/T_c)^4)^2} $ which can be expressed as
\be
\frac{S_3}{T} \approx
\frac{2.95 \  N^{7/2}} {\left|\epsilon (\frac{\delta T_1}{2}\mp
v_1^2\sqrt{-{\delta T_1 \over v_1^2}+\epsilon } )\right|^{3/4}}   \times \frac{(T_c/T)}{({1-(T/T_c)^4})^2  }
\label{S3overT}
\ee
In this formula, we have neglected the unknown barrier contribution
which could change the result by a factor of order unity.
It turns out the transition is dominated by the thick wall regime even
without additional suppression so that this formula is adequate.


The $T$-dependent factor in the right of eq. (\ref{S3overT}) blows up as $T$ approaches $T_c$. As $T$ slightly decreases, it becomes of order one, reaches a minimum at $T=T_c/3^{1/2}$ and grows again. Clearly, the nucleation condition $S_3/T \lsim  140$ is satisfied when this factor is near its minimum, for which it is $\sim {\cal O}(1)$.
This typically takes place when $T$ is between $T_c/3^{1/2}$ and 0.9 $T_c$.

From eq.(\ref{S3overT}), it is clear that there is a constraint that $N$
 not be too large (or equivalently, the AdS curvature scale not be too small).
 This follows from  the large entropy in the
 AdS-Schwarschild phase, which grows as $N^2$, whereas the RS phase entropy does
 not. This means that the phase transition is entropically disfavored for large
 $N$. We can understand the $N$-dependence in the action as
arising from the radion-normalization and from the shallowness of the
 potential.  This latter $N$-dependence comes from the size of the bubbles, which grows with $N$, and would not be present if $v_1$ was as large as
 $N$. However perturbativity of the model requires $v_1<N$. We will comment
 further on this point below.

\begin{figure}[!htb]
\begin{center}
\includegraphics[height=5.2cm,width=8.1cm]{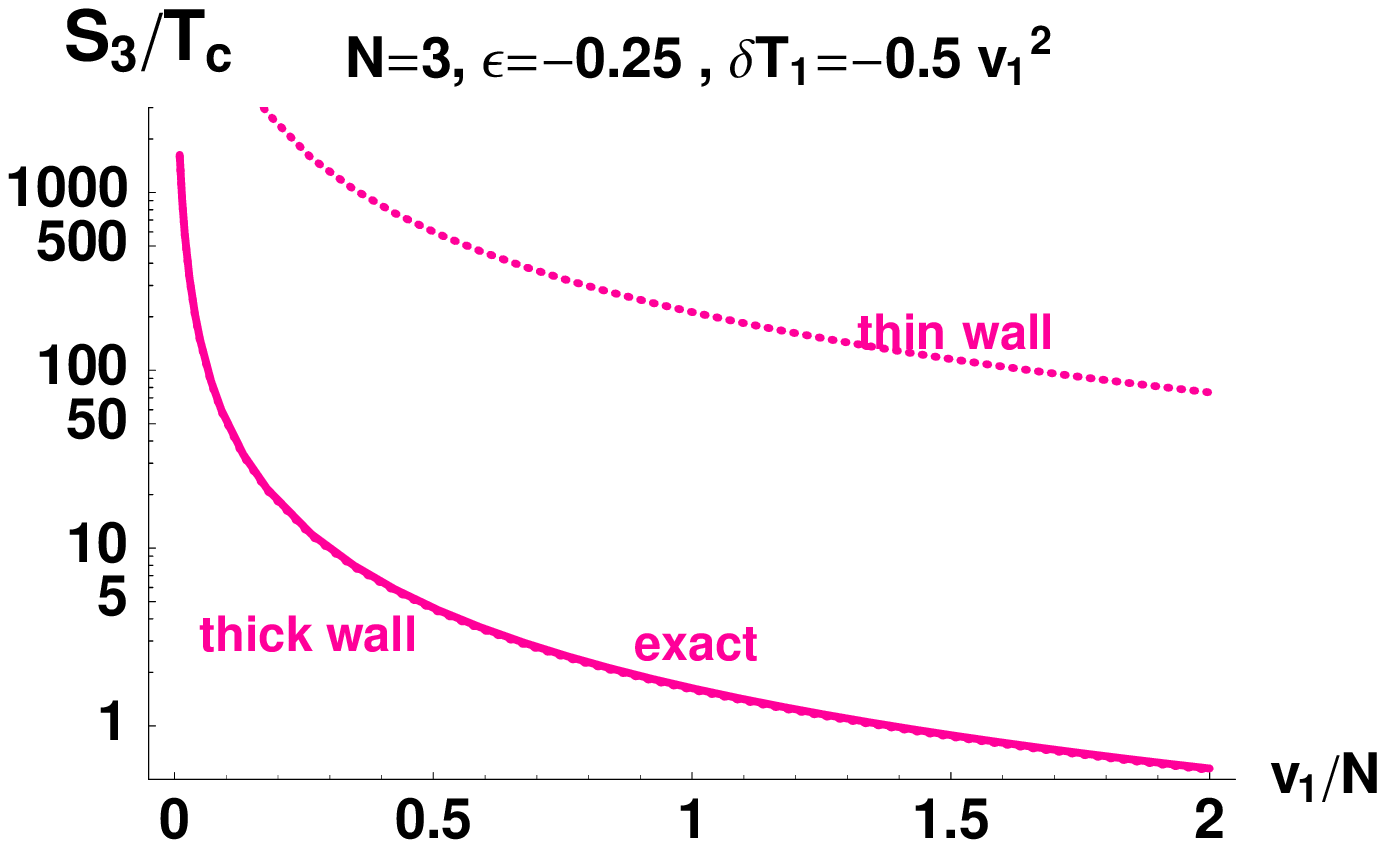}
\includegraphics[height=5.2cm,width=8.1cm]{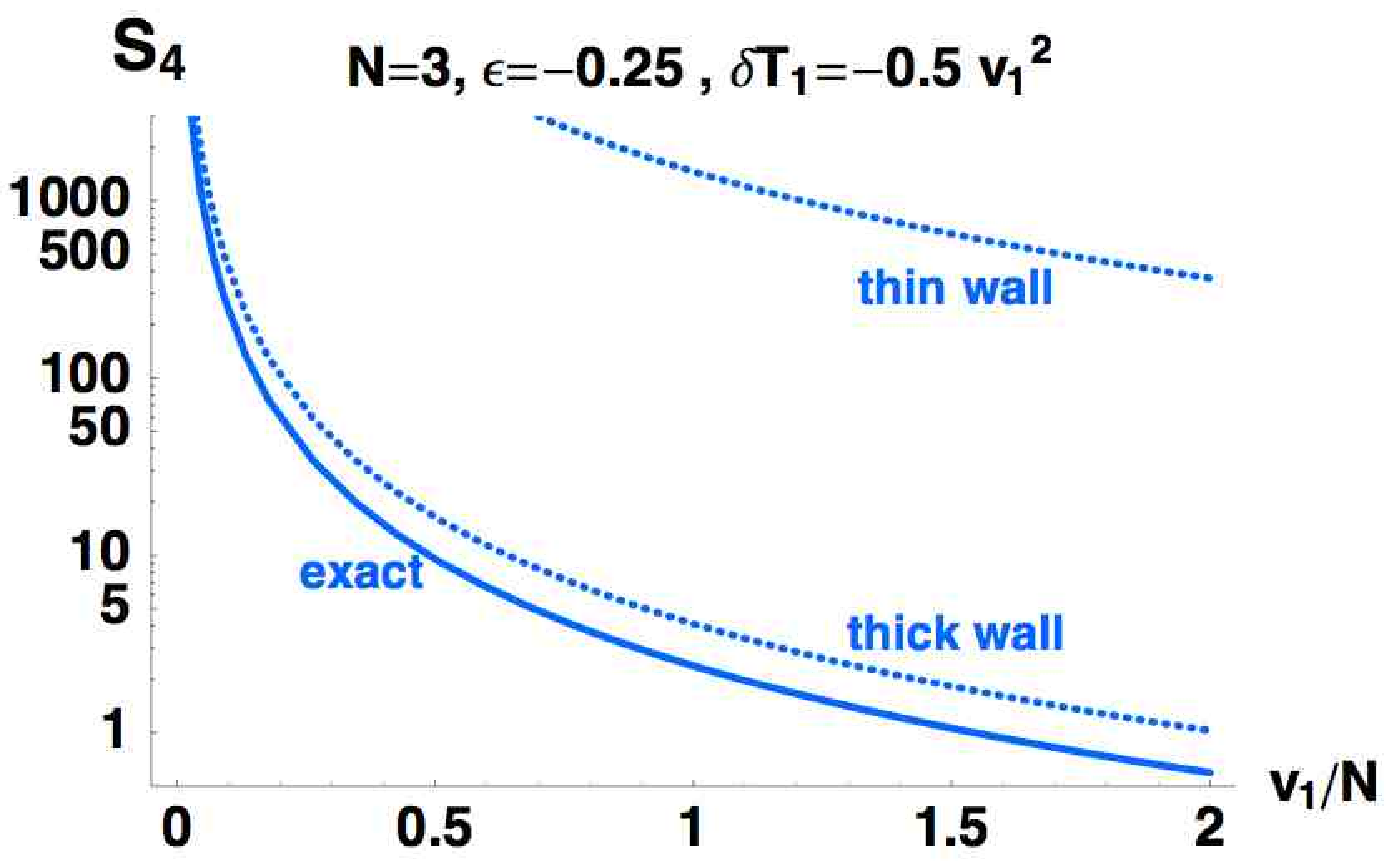}
\includegraphics[height=5.2cm,width=8.1cm]{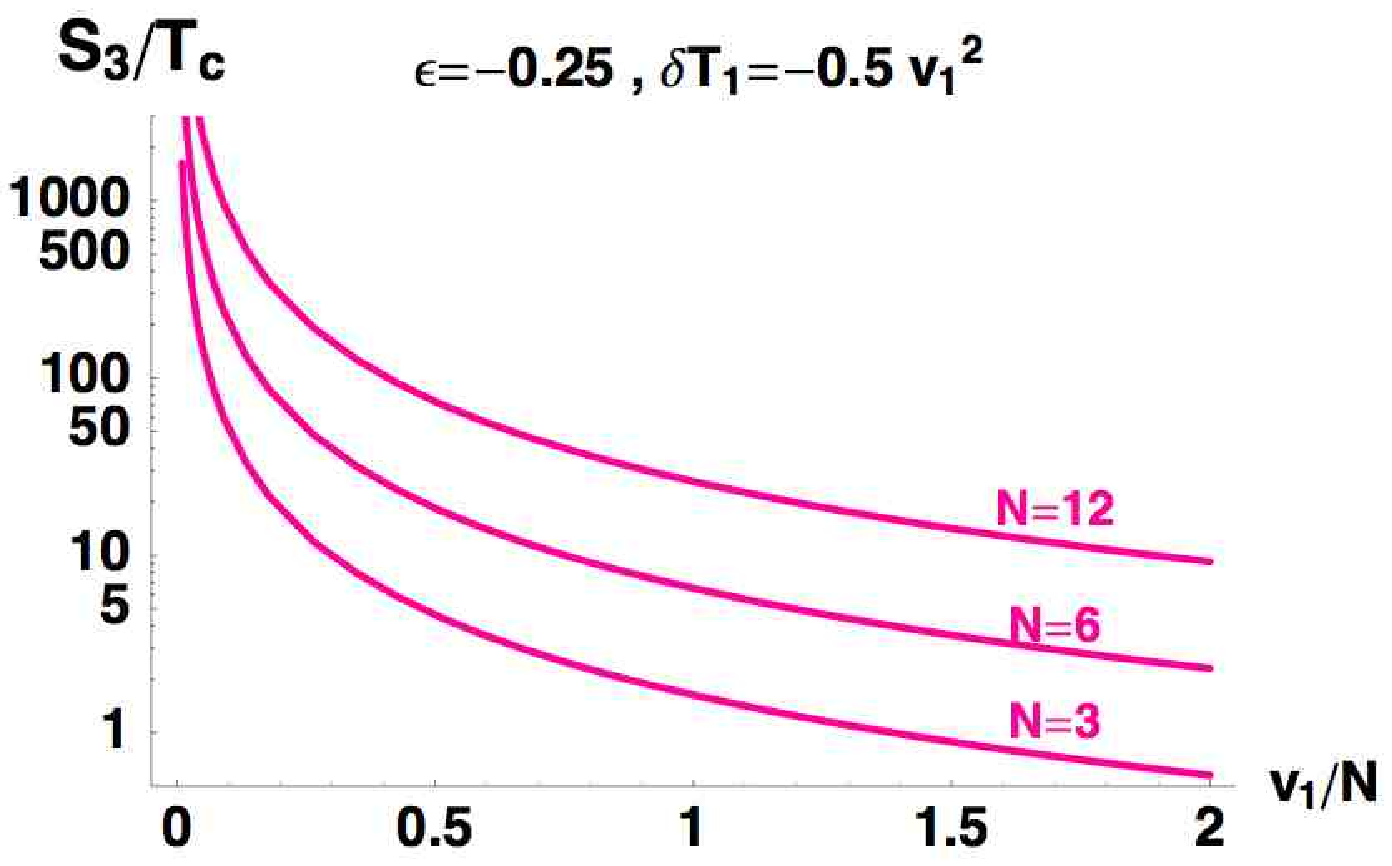}
\includegraphics[height=5.2cm,width=8.1cm]{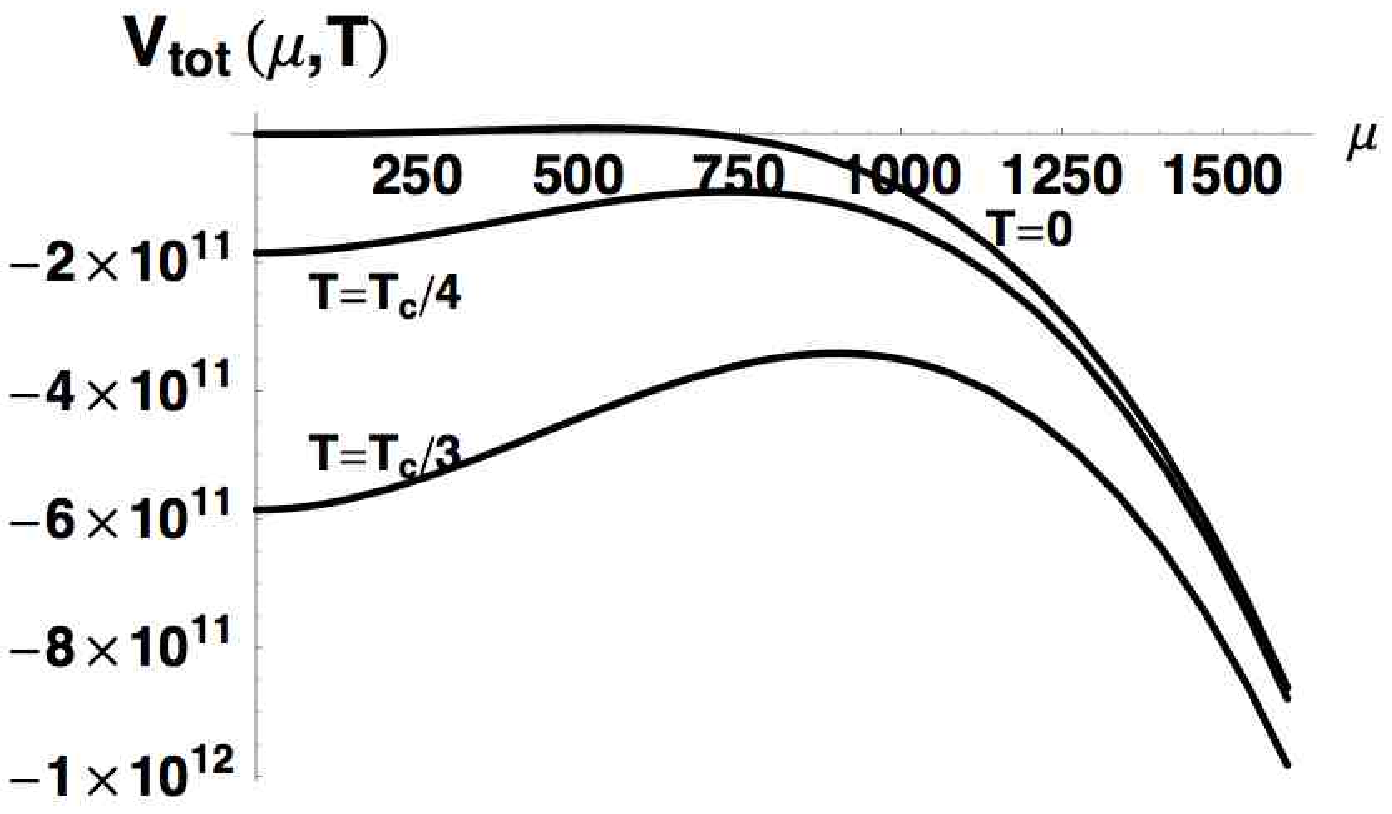}
\caption{Top: comparison between thick wall, thin wall and exact solutions at fixed $\epsilon$ and $\delta T_1$; bottom left: exact results for different values of $N$. bottom right: Typical evolution of the radion potential with temperature. The height of the barrier falls off as $T$ goes down. For $T$ below $T_c/2$, it is a very good approximation to use the zero temperature potential to compute the bounce.}
\label{comp_v1}
\end{center}
\end{figure}

  We have ignored the $T$-dependent corrections to the radion potential so far. We can model the radion potential at finite temperature just assuming that the mass of the CFT degrees of freedom is proportional to the vev of $\mu$, $m= g \mu$, according to the general formula:
 \be
 \Delta V(\mu,T)=\sum_{F} \frac{g_{F}T^4}{2\pi^2}
\sum_n \frac{(-1)^n}{n^2}(\frac{m_{F}}{T})^2K_2(\frac{nm_{F}}{T})
-\sum_{B} \frac{g_{B}T^4}{2\pi^2}
\sum_n \frac{1}{n^2}(\frac{m_{B}}{T})^2K_2(\frac{nm_{B}}{T})
 \ee
This is the 1-loop  thermal corrections to the radion potential taking into account the interactions of the radion with the $N^2$ CFT degrees of freedom. $\sum_{B/F}g_{B/F}=45 N^2/8 \pi^2$.
 At $\mu=0$, this reproduces the $T^4$ radiation energy density, but as soon as $\mu$ is of order $T$, the CFT degrees of freedom  are massive and the $T$-dependent piece falls off.
As illustrated on the fourth plot of Fig.~\ref{comp_v1}, at $T$ close to $T_c$, the barrier is big and prevents tunneling.
But as $T$ goes below $T_c/2$, the barrier becomes much smaller than the energy difference  between the two minima and we tend toward the zero temperature radion potential.   The transition does not take place until the temperature is low enough so that we are in the thick wall approximation.
This trend is general, even though detailed predictions will depend on the precise coefficient  $g$ between the mass of the CFT degrees of freedom and the radion. We took $g=1$ in the right bottom plot of Fig.~\ref{comp_v1}.

Thin-wall bubbles generally yield too big an action for nucleation and we now show that thick-wall
bubbles   dominate. As we have just argued, it is a good approximation to use the zero-temperature radion potential to estimate the tunneling action.
The action for thick-walled bubbles is $S_3\approx 2 \pi R \mu^2 -4\pi R^3|\Delta V|/3$ \cite{Anderson:1991zb} and its minimal value $S_3=8\pi R^3|\Delta V|/3$ associated with $R=\mu \sqrt{3N^2/2\pi^2} /\sqrt{2|\Delta V|}$ leads to
\be
\label{S3action}
 S_3\approx({4\pi}/{3})
 ({\mu}^3/{ \sqrt{2 | \Delta V_{\pm}| }}) \left({3N^2}/{2\pi^2}\right)^{3/2}
 \ee
which is, assuming again that at nucleation $\mu=\mu_{\mbox{\tiny TeV}} $:
\be
\label{S3actionapprox}
 S_3\approx({4\pi}/{3})
 ({{{\mu}_{\mbox{\tiny TeV}}}}^3/{ \sqrt{2 |V_{\pm}|}})\left({3N^2}/{2\pi^2}\right)^{3/2}  / (1-(T/T_c)^4)^2
 \ee
This action is  smaller by an overall numerical factor (sixteen) than the thin wall  formula.

We also consider the possibility of $O(4)$-symmetric vacuum bubbles. These can be relevant only if the nucleation does not take place via thermal bubbles.
The thin-wall action in this case is  \cite{Linde:1981zj}
$S_4=\ 27 \pi^2 S_1^4\times (3N^2/(2\pi^2))^2/ 2 |\Delta V_{\pm}|^3$
whereas the thick-wall action is $S_4=\pi^2 R^2 \mu^2-\pi^2R^4|\Delta V|/2$, which is minimal for $R=\mu \sqrt{3N^2/2\pi^2}/\sqrt{|\Delta V|}$
leading to
\begin{equation}
\label{S4action}
S_4=\pi^2 \mu^4(3N^2/2\pi^2)^2/(2|\Delta V_{\pm}|).
\end{equation}
and if  $\mu=\mu_{\mbox{\tiny TeV}} $:
\be
\label{S4actionapprox}
S_4=\pi^2 \mu_{\mbox{\tiny TeV}}^4(3N^2/2\pi^2)^2/(2| V_{\pm}|(1-a^4)).
\ee
The ratio of the thick-wall to thin-wall action is $N$-independent and for $\mu={\mu}_{\mbox{\tiny TeV}}$, it scales as:
\be
\frac{S_4}{S_3/T}=\frac{3\sqrt{3}}{8^{3/4}\sqrt{\pi}} \ \frac{\frac{T}{T_c}}{\sqrt{1-(\frac{T}{T_c})^4}}\frac{(2+\epsilon)^{1/4}}{\sqrt{b}\epsilon^{1/4}[-c\mp\sqrt{4c+\epsilon(4+\epsilon)}]^{1/4}}
\ee
It will thus be more favorable to tunnel via $O(4)$-symmetric bubbles if $T/T_c$ is small, $\epsilon$ is large and $b=v_1/N$ and $c=|\delta T_1|/v_1^2$ are large. Using the  $O(4)$ rather than the $O(3)$ symmetric solution corresponds to $T_n<(2R)^{-1}$.

The value of  $\mu$ to be used in eq. (\ref{S3action}) and (\ref{S4action}) is actually not $\mu_{\mbox{\tiny TeV}}$ but the value at the time of the tunneling which is typically smaller than $\mu_{\mbox{\tiny TeV}}$. That is because  the value of the field that minimizes the action is not exactly  the value of the field at the potential minimum \cite{Anderson:1991zb}. The field tunnels to a value near the minimum and after tunneling rolls to the minimum.
We first evaluated the nucleation action assuming that $ \mu= \mu_{\mbox{\tiny TeV}}$ at the time of nucleation, to get some analytical dependence on parameters. To derive our results, we do not apply formula (\ref{S3actionapprox}) and (\ref{S4actionapprox}) where the radion value is taken at the minimum of the potential. We search for  the value of the radion field $\mu$  minimizing the bubble free energy, which is, for the $O(3)$ action,
\be
\label{mu3}
{\mu}=6\frac{|\Delta V_{\pm}|}{|\Delta V_{\pm}|'}
\ee
and for the $O(4)$ action
\be
\label{mu4}
{\mu}=4\frac{|\Delta V_{\pm}|}{|\Delta V_{\pm}|'}
\ee
where
\be
|\Delta V_{\pm}|=V_{\pm}[a^4-\frac{V_{\pm}(\mu)}{V_{\pm}}] \ \ \ , \ \ \ a=\frac{T_n}{T_c}
\ee
and
\be
V_{\pm}=\mu_{\mbox{\tiny TeV}}^4 \ v_1^2 \ \frac{\epsilon}{2+
\epsilon}[c\pm\sqrt{4c+\epsilon(4+\epsilon)}]
\ee
is the value of the potential at the minimum of the potential and we remind that $c= |\delta T_1/v_1^2|$.
We re-express everything in terms of the variable
\be
Y\equiv \left(\frac{\mu}{\mu_{\mbox{\tiny TeV}}}\right)^{\epsilon}
\ee
Defining
\be
X_{\pm}\equiv \frac{1+\frac{\epsilon}{4}\pm
\frac{1}{2}\sqrt{c+\epsilon(1+\frac{\epsilon}{4})}}{1+\frac{\epsilon}{2}}
\ee
we get
\be
V_{\pm}(Y)=\mu_{\mbox{\tiny TeV}}^4 Y^{\frac{4}{\epsilon}} \ v_1^2 \ [(4+2 \epsilon)(1-X_{\pm}Y)^2-\epsilon-c]
\ee
Eq (\ref{mu3}) and (\ref{mu4}) can be rewritten as
\be
2(2-\epsilon) X_{\pm}^2 Y^2 -8  X_{\pm}Y + 4 - \epsilon-c -3 a^4 Y^{-\frac{4}{\epsilon}}\epsilon (2-\epsilon)[-c\mp
\sqrt{4c+\epsilon(4+\epsilon)}]=0
\label{eqY3}
\ee
\be
(2+\epsilon) (4+2\epsilon)X_{\pm} Y (1-X_{\pm}Y) -2 a^4 Y^{-\frac{4}{\epsilon}} [-c\mp
\sqrt{4c+\epsilon(4+\epsilon)}]=0
\label{eqY4}
\ee
\begin{figure}[!htb]
\begin{center}
\includegraphics[height=5.cm,width=8.1cm]{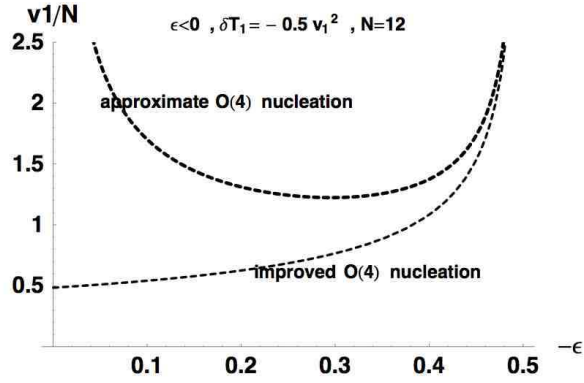}
\caption{Lines delimiting the region in $(\epsilon, v_1)$ parameter space where the phase transition completes (we have to be above the dashed line for nucleation to take place). The upper line comes from using the approximate eq (\ref{S4actionapprox}) while the lower line results from solving eq.  (\ref{mu4}).}
 \label{comparison}
\end{center}
\end{figure}
 We  then look for the regions in $(\epsilon, v_1)$ plane where the phase transition completes. We evaluate $S_3/T$ and $S_4$ at the $Y$ solution of eq  (\ref{eqY3}) and (\ref{eqY4}) and follow  their evolution with temperature. The nucleation temperature is defined by the condition $S_3/T=140$ or $S_4=140$.  Whether  $S_3/T$  or $S_4$ first reaches this critical value  determines whether $O(3)$ or $O(4)$ bubbles are nucleated. Note that the $O(4)$ nucleation region is simply given by $v_1/N>3N/(4\pi\sqrt{70(c+\epsilon)})$ where the right-hand side corresponds to the limit $T_n \rightarrow 0$.
To illustrate how much difference this procedure makes, we compare the approximate solution (using $ \mu= \mu_{\mbox{\tiny TeV}}$) and the accurate one in Fig.~\ref{comparison} for  some particular choice of parameters  in the $O(4)$ case.  We get the same kind of discrepancy for $O(3)$.  The biggest effect is for small $\epsilon$ in which case the potential at the minimum is very shallow. Consequently, it does not cost much potential energy to tunnel to the wrong place, namely to $\mu < {\mu}_{\mbox{\tiny TeV}}$.

\begin{figure}[!htb]
\begin{center}
\includegraphics[height=6.4cm,width=8.1cm]{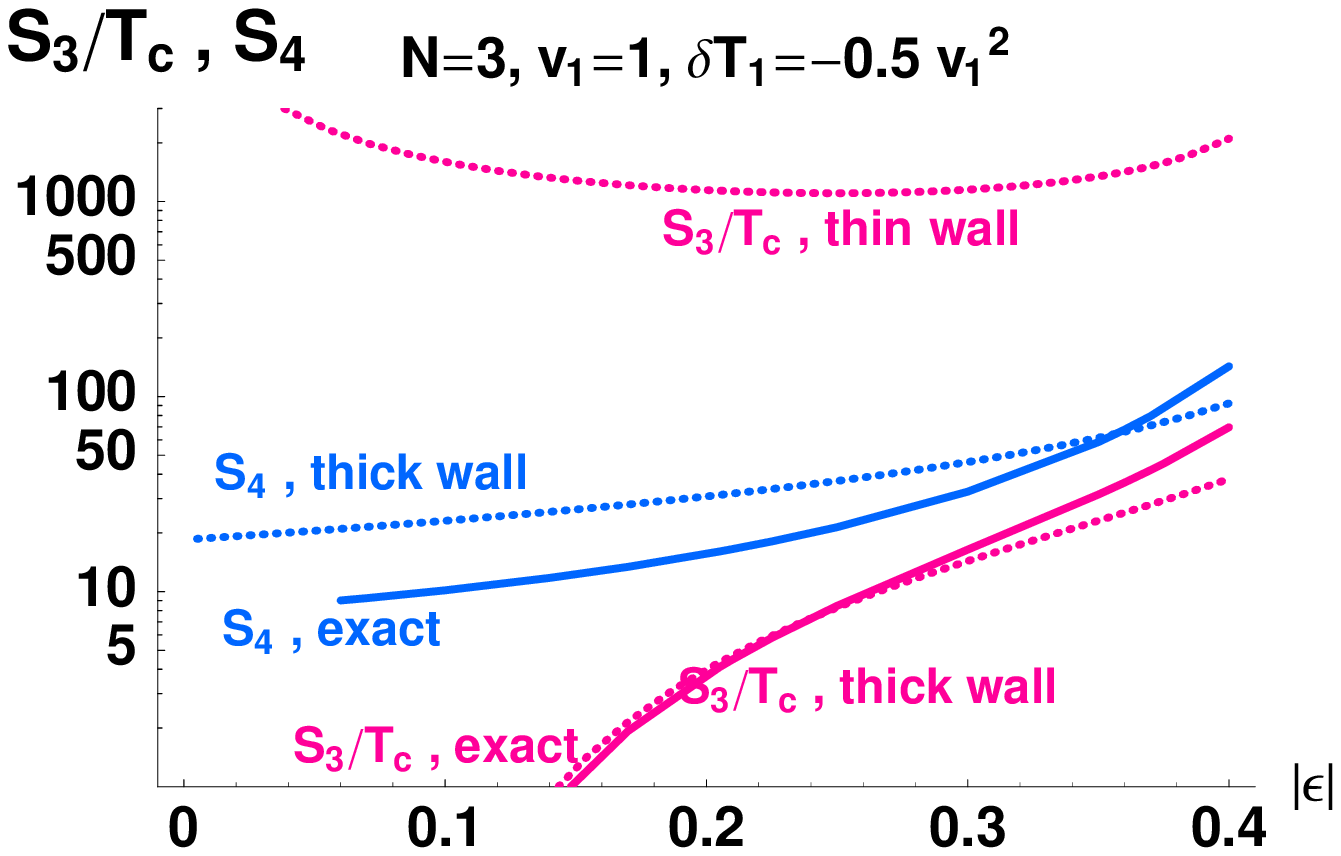}
\includegraphics[height=6.4cm,width=8.1cm]{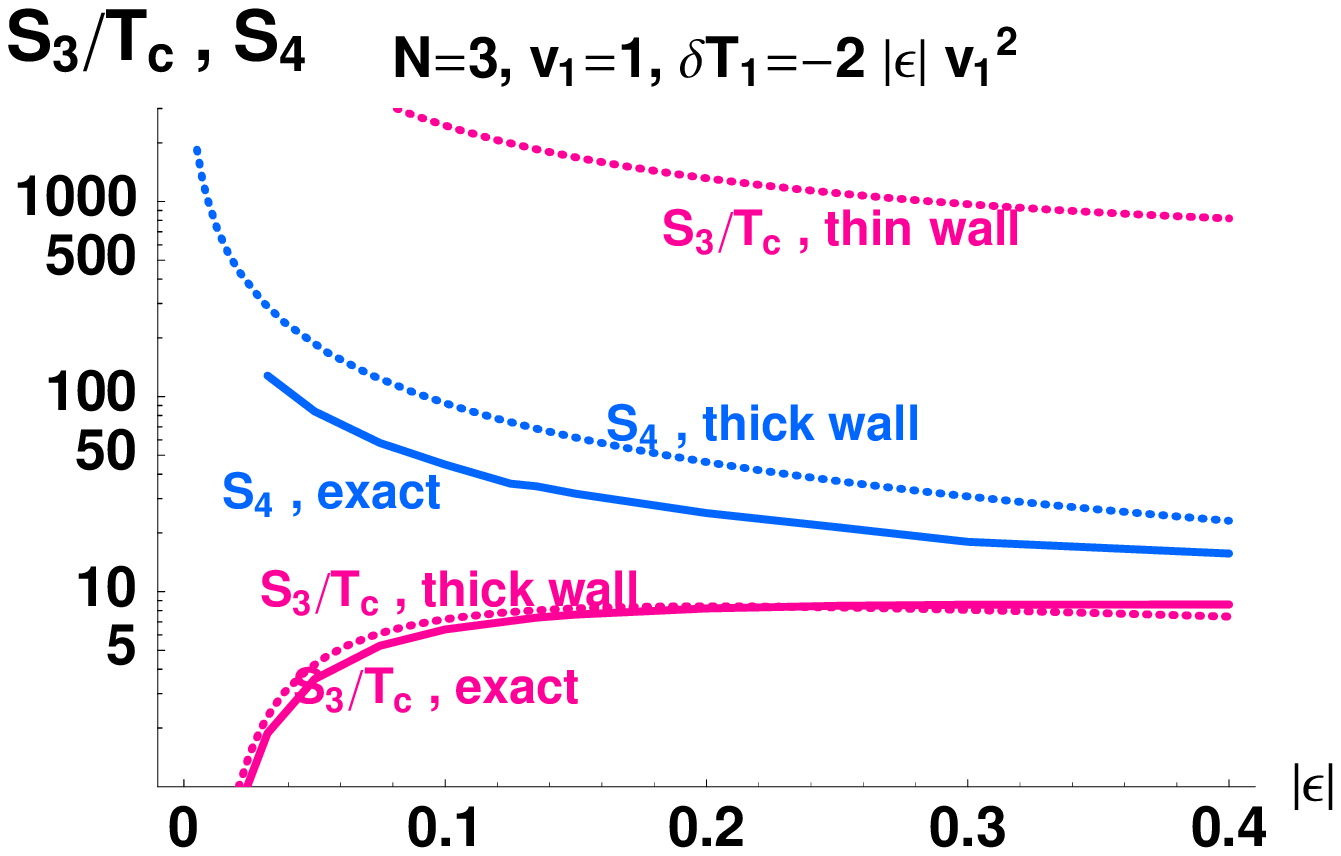}
\caption{Comparison of the thin and thick wall approximations (dotted lines) with the exact solutions obtained by solving for the bounce numerically (solid lines).}
\label{comp_eps}
\end{center}
\end{figure}

We have checked in various regions of parameter space that the thick wall approximation
gives an answer that is indeed close to the exact bounce computed numerically. This is illustrated in Fig.~\ref{comp_eps} and Fig.~\ref{comp_v1}.

Finally, we note that there exists the possibility of an inflationary phase before the phase transition occurs. This is due to the vacuum energy
$-V_{\pm}$ in the AdS-Schwarschild phase if we set the cosmological constant to 0 in the RS phase. Inflation could in principle start when the temperature is of order $T_c$.
However, even if inflation were to occur, the inflationary epoch would last only a very short time as the number of e-foldings would be only $\ln ({T_c}/{T_n})$. Therefore, even if the universe inflates above the nucleation temperature, the transition would typically complete in  less than a single e-folding. This is true even if the universe settles into the wrong ($\mu=0$) minimum at early times. In fact, because we neglected the AdS-S contribution to the bounce, the action associated with tunneling from this false minimum would be precisely the same as what we've calculated.

To summarize this section,  we have extended the Creminelli et al calculation to the $\epsilon<0$ and
 $\delta T_1 \neq 0$ cases. In addition, we considered the nucleation of thick-walled bubbles rather than thin-walled ones, which applies in the regime of large supercooling.
We computed the free energy of a critical bubble by searching for the value of the radion field that minimizes the action, rather than assuming that the radion value sits at the minimum of the potential at the time of nucleation. 
Also relevant in the regime of large supercooling is the nucleation of $O(4)$ symmetric bubbles. These effects improve the nucleation probability.

 \section{Gravitational wave signal}

 We  can now determine $\alpha$ and $\beta$ and hence the gravitational
 wave signal. We repeat that we are
 working under the assumption that the phase transition is very strong (as
 will be justified below), which is why predictions can be given as functions of $\alpha$ and $\beta/H_*$ only.
The signal obviously grows with $\alpha$ as the more latent heat that is
 released, the more measurable is the phase transition.
The quantity $\alpha$ can be estimated as
\be
\alpha =\frac{|\Delta V_{\pm}|}{\pi^2N^2T^4/8}=\frac{1}{a^4}\frac{V_{\pm}(Y)}{V_{\pm}}-1
\ee
To get a feeling for the results, we will first assume that at nucleation $\mu={\mu}_{\mbox{\tiny TeV}}$, in which case
 $\Delta
 V_{\pm} =-\pi^2 N^2 T_c^4(1-(T_n/T_c)^4)/8$.
 Hence we can determine $\alpha$ simply in terms of the ratio of the nucleation to the critical temperature as
\be
\alpha =(T_c^4-T_n^4)/T_n^4
\label{alphaformula}
\ee
The size of the signal also increases with the duration of the phase
 transition $\beta^{-1}$
 given in terms of $a={T_n}/{T_c}$ by:
\bea
\label{betaformula}
 \frac{\beta}{H}&=&  \left\{ \begin{array}{ll}
 {S_3(T_n)}/{T_n}  \times    \frac{3a^4-1}{1-a^4  } \approx 140 \times \frac{3a^4-1}{1-a^4  } \ \ \ \mbox{for $O(3)$ solution}\\
 {S_4(T_n)}  \times    \frac{4a^4}{1-a^4  } \approx 140 \times \frac{4a^4}{1-a^4  } \ \ \ \mbox{for $O(4)$ solution}\\
	\end{array} \right.\\
	\nonumber
\eea
\begin{figure}[!htb]
\begin{center}
\includegraphics[height=5.5cm,width=8.1cm]{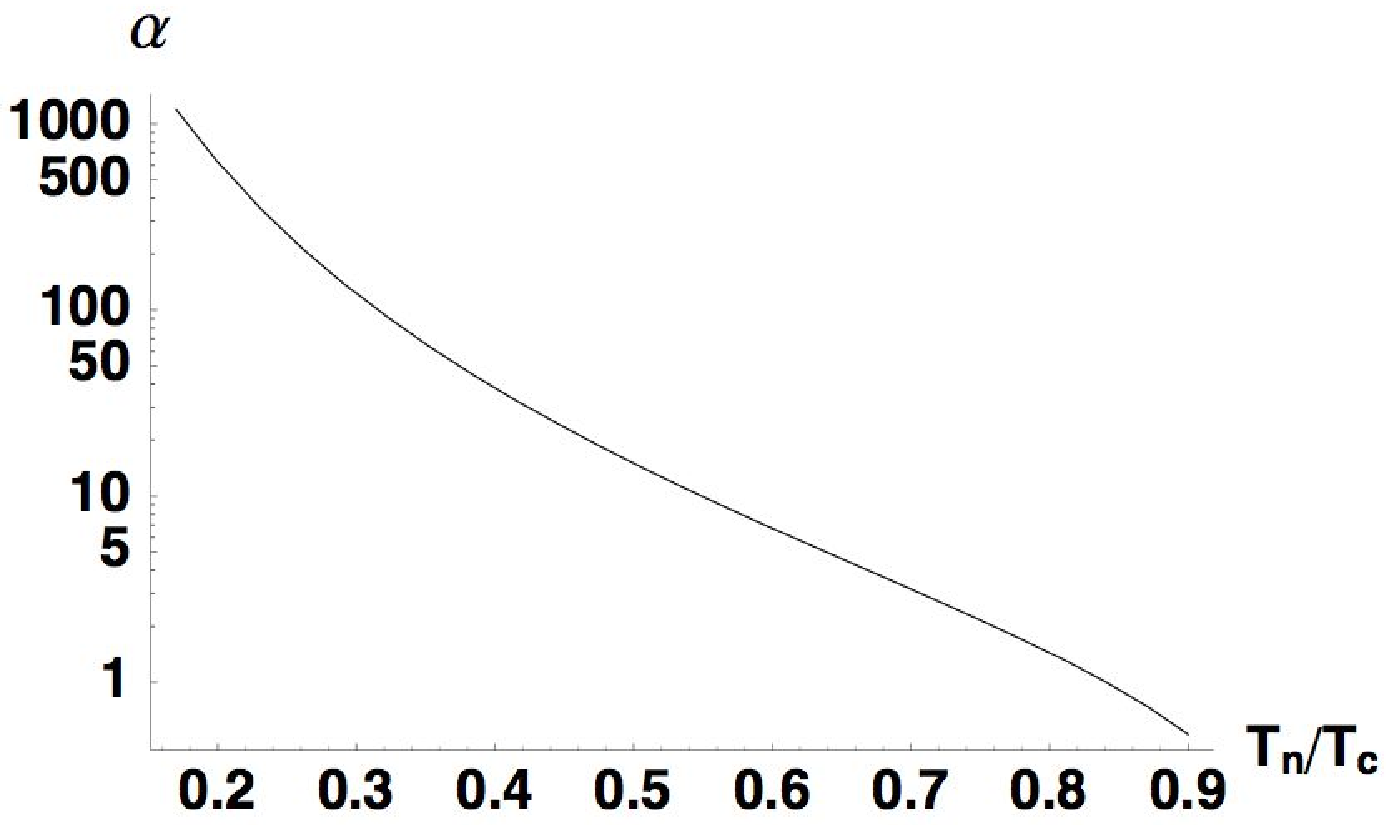}
\includegraphics[height=5.5cm,width=8.1cm]{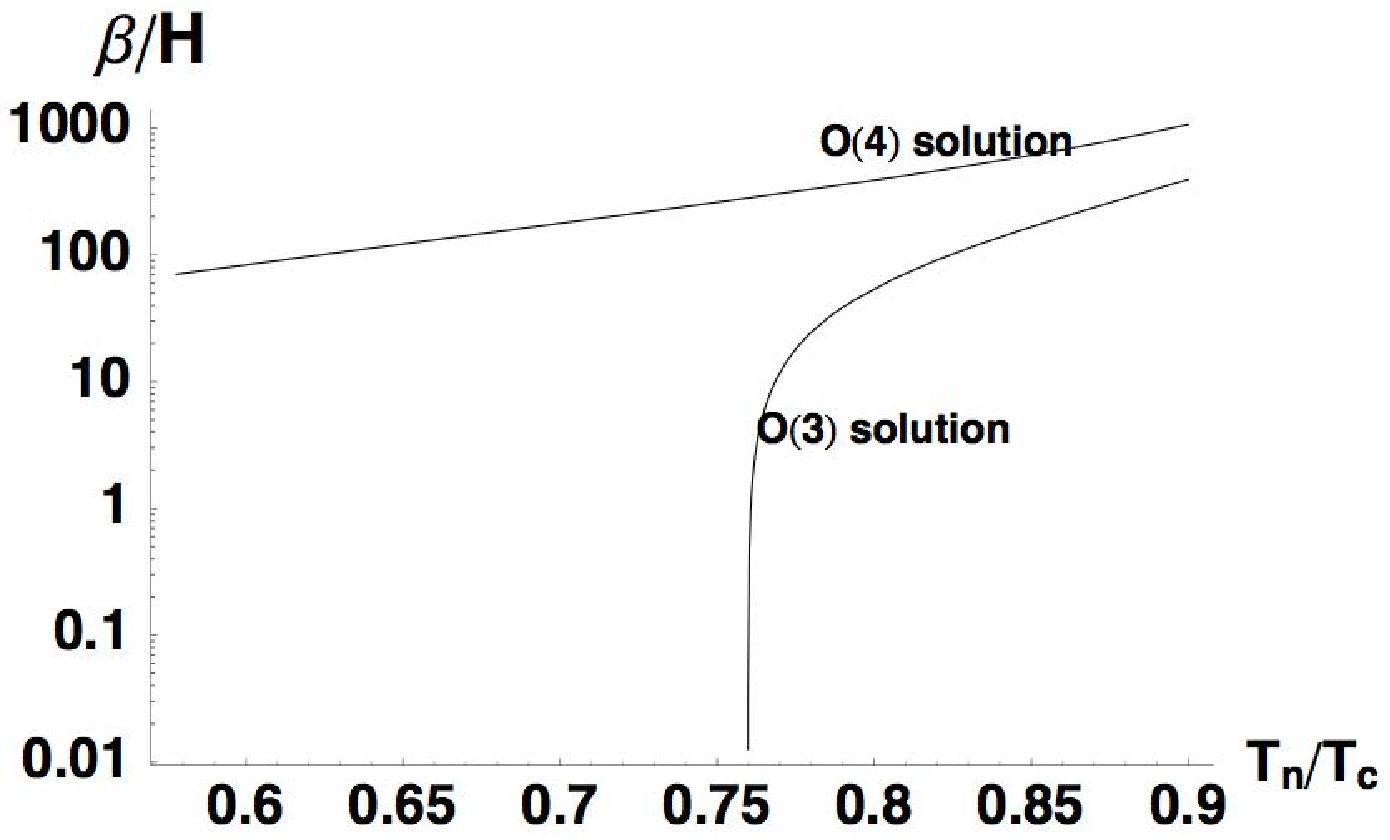}
\caption{$\alpha$ is the ratio of the latent heat to the radiation energy density in the CFT phase at the time of nucleation, given by Eq.~(\ref{alphaformula}). It increases as the ratio of the nucleation temperature $T_n$ to the critical temperature $T_c$ decreases. Second plot is $\beta/H $ from Eq.~(\ref{betaformula}) where $\beta^{-1}$ can be understood as the duration of the phase transition. The amplitude of the gravitational wave signal increases with $\alpha$ and decreases as $(\beta/H)^{-2} $.}
 \label{alphaversusN}
\end{center}
\end{figure}
 \begin{figure}[!htb]
\begin{center}
\includegraphics[height=4.75cm,width=8.1cm]{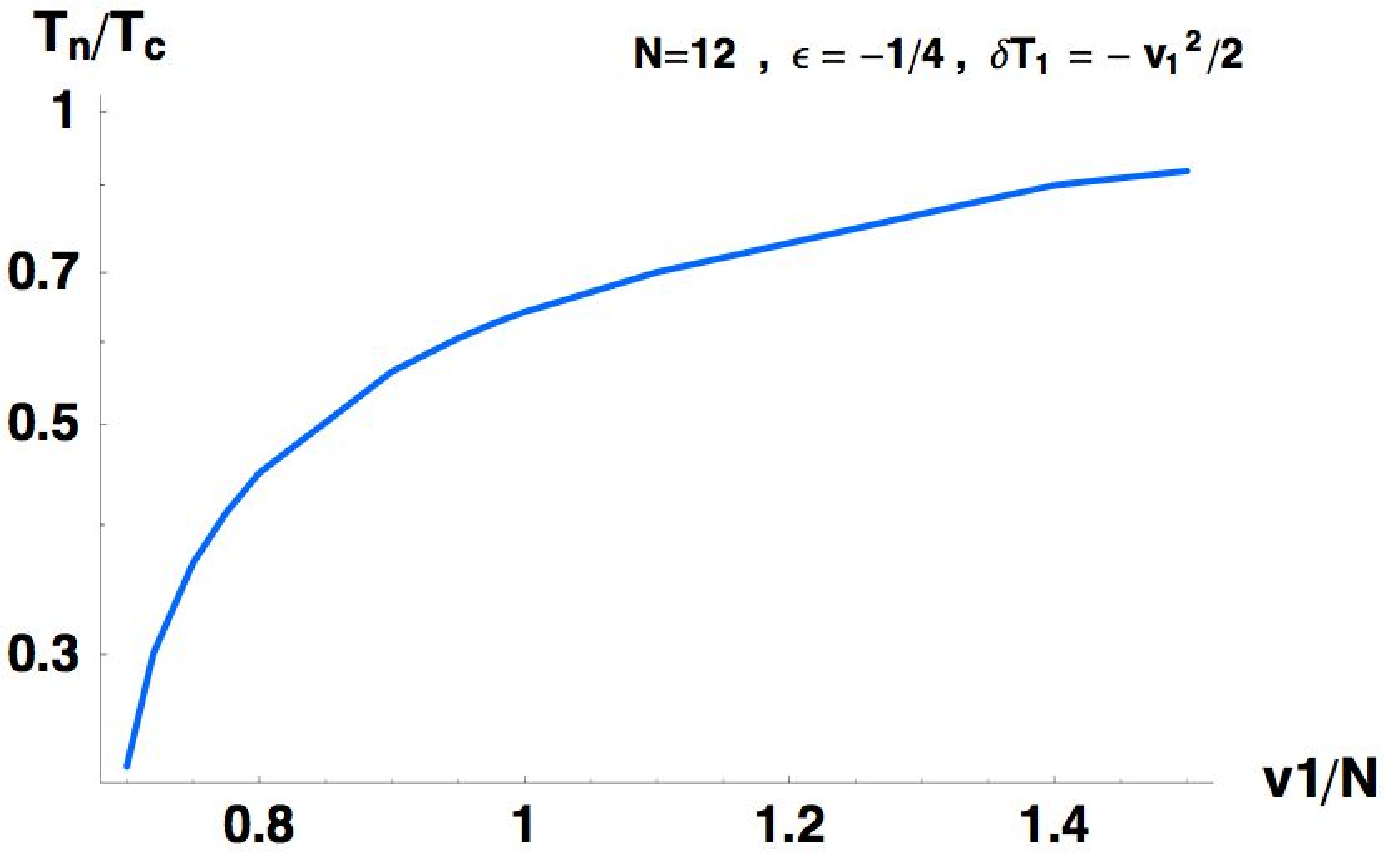}
\includegraphics[height=4.75cm,width=8.1cm]{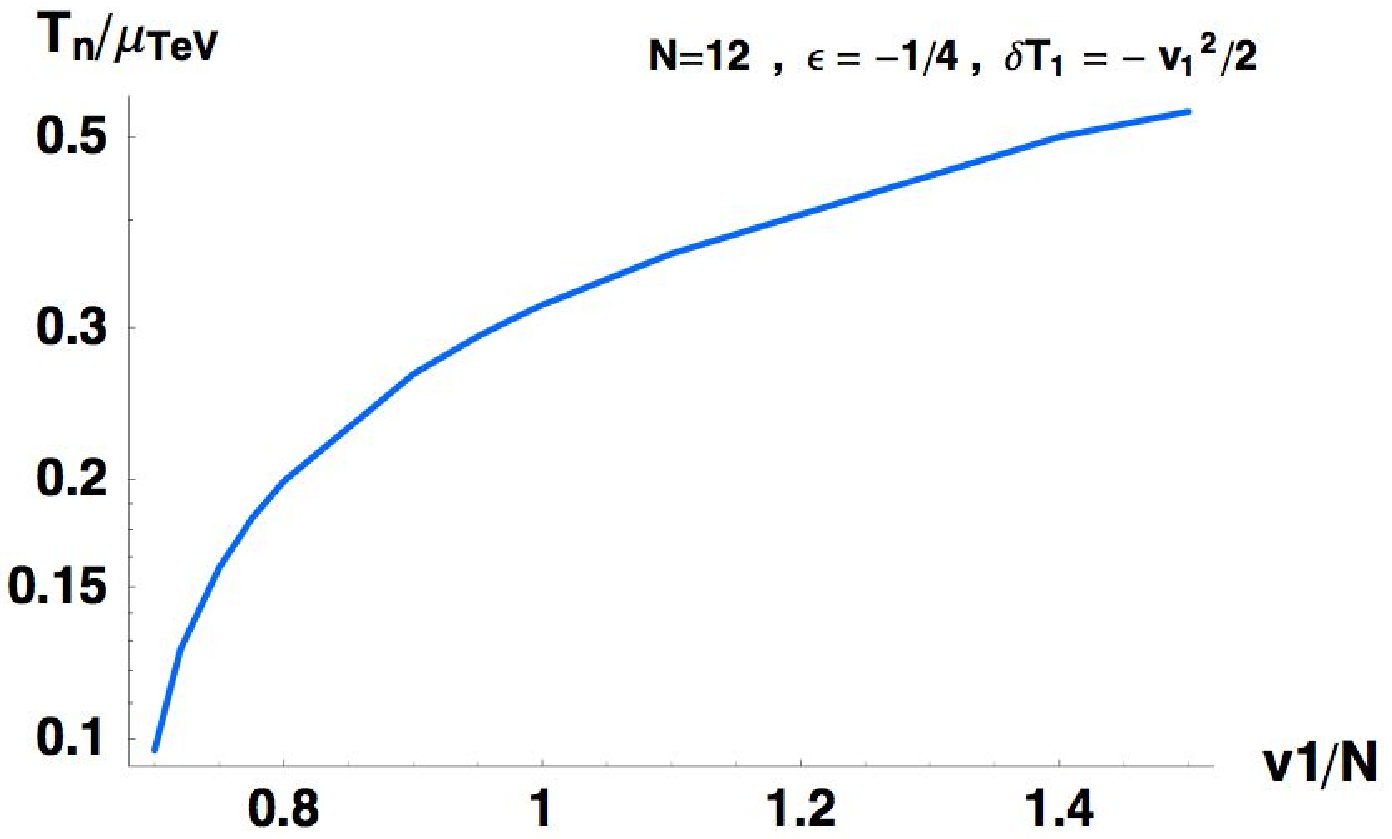}
\includegraphics[height=4.75cm,width=8.1cm]{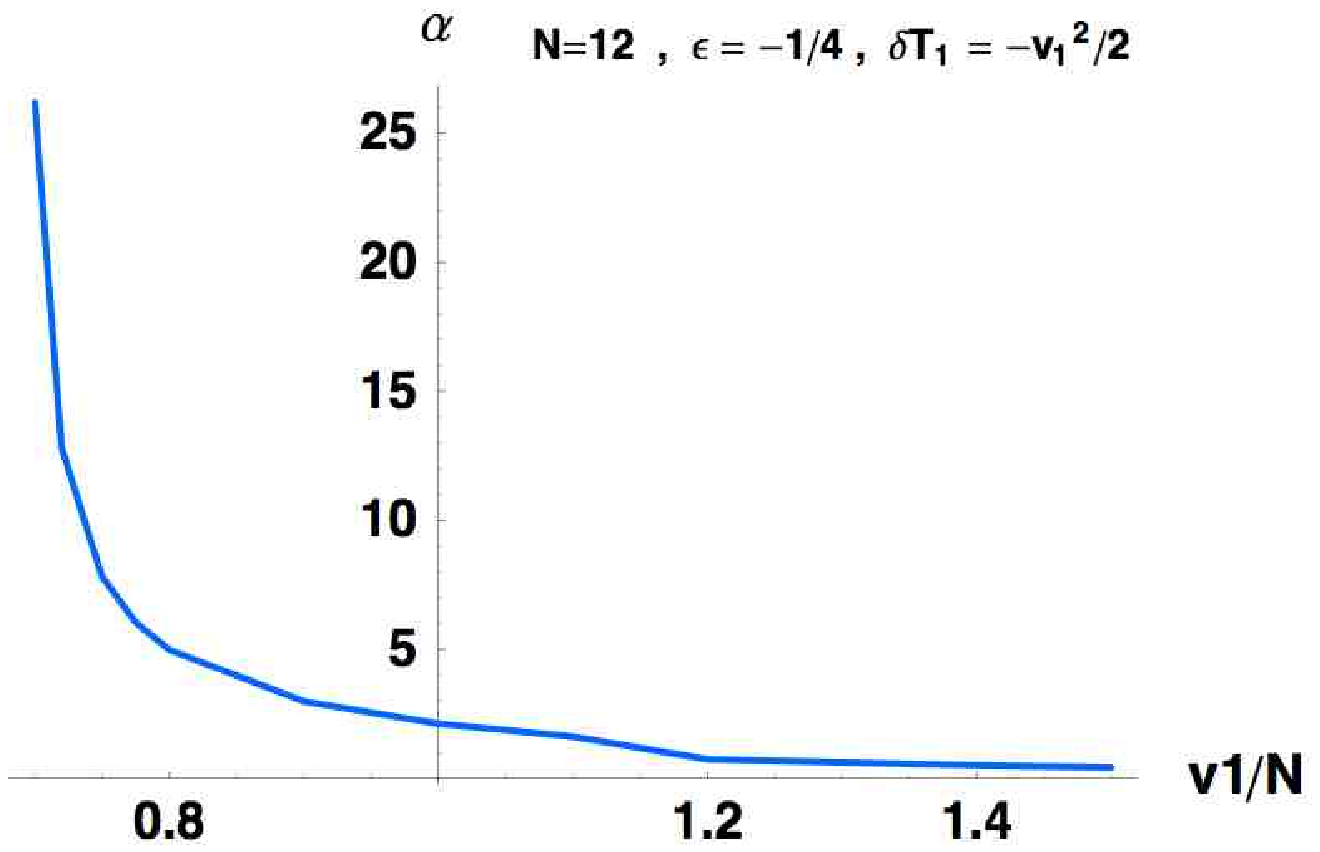}
\includegraphics[height=4.75cm,width=8.1cm]{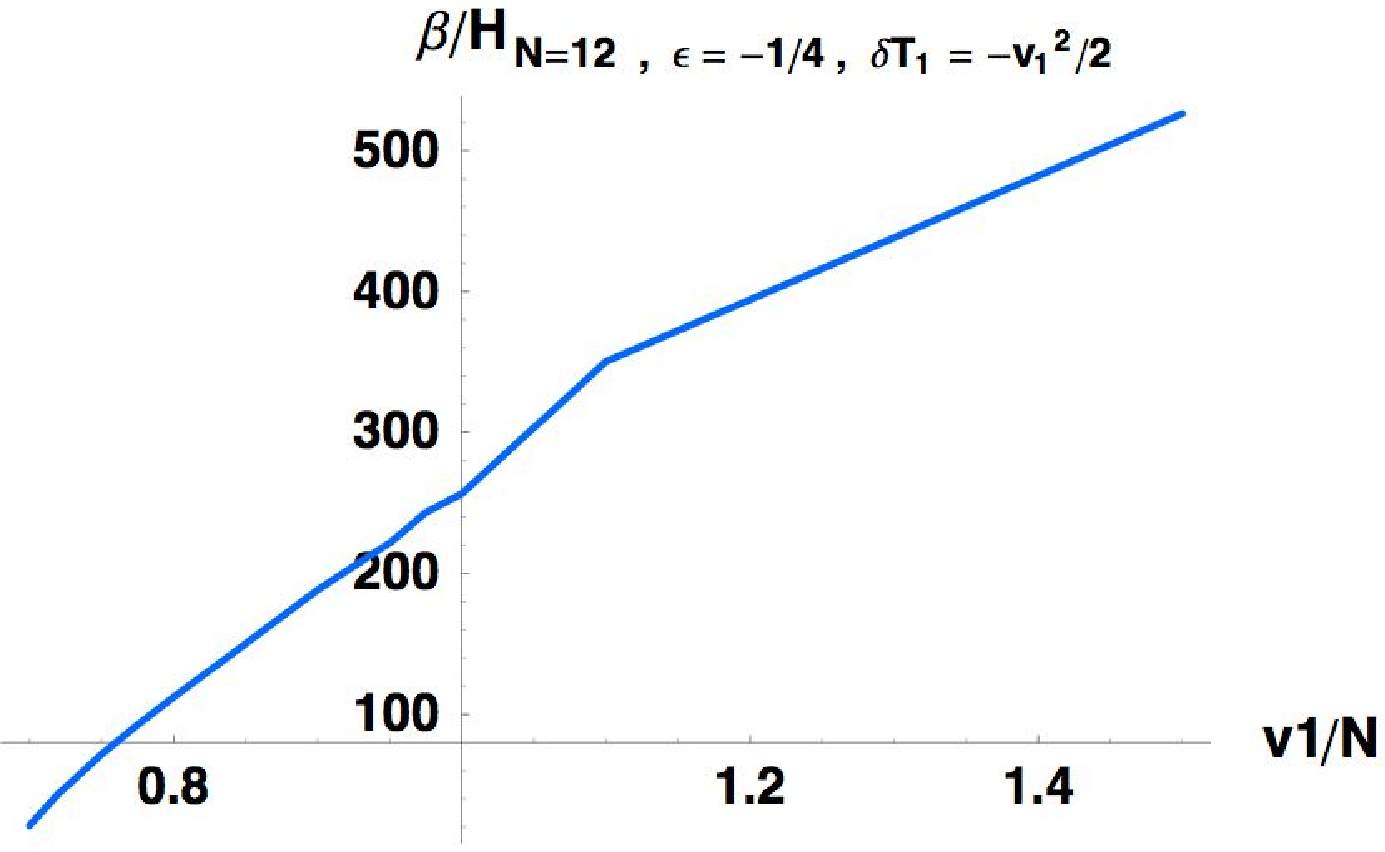}
\includegraphics[height=4.75cm,width=8.1cm]{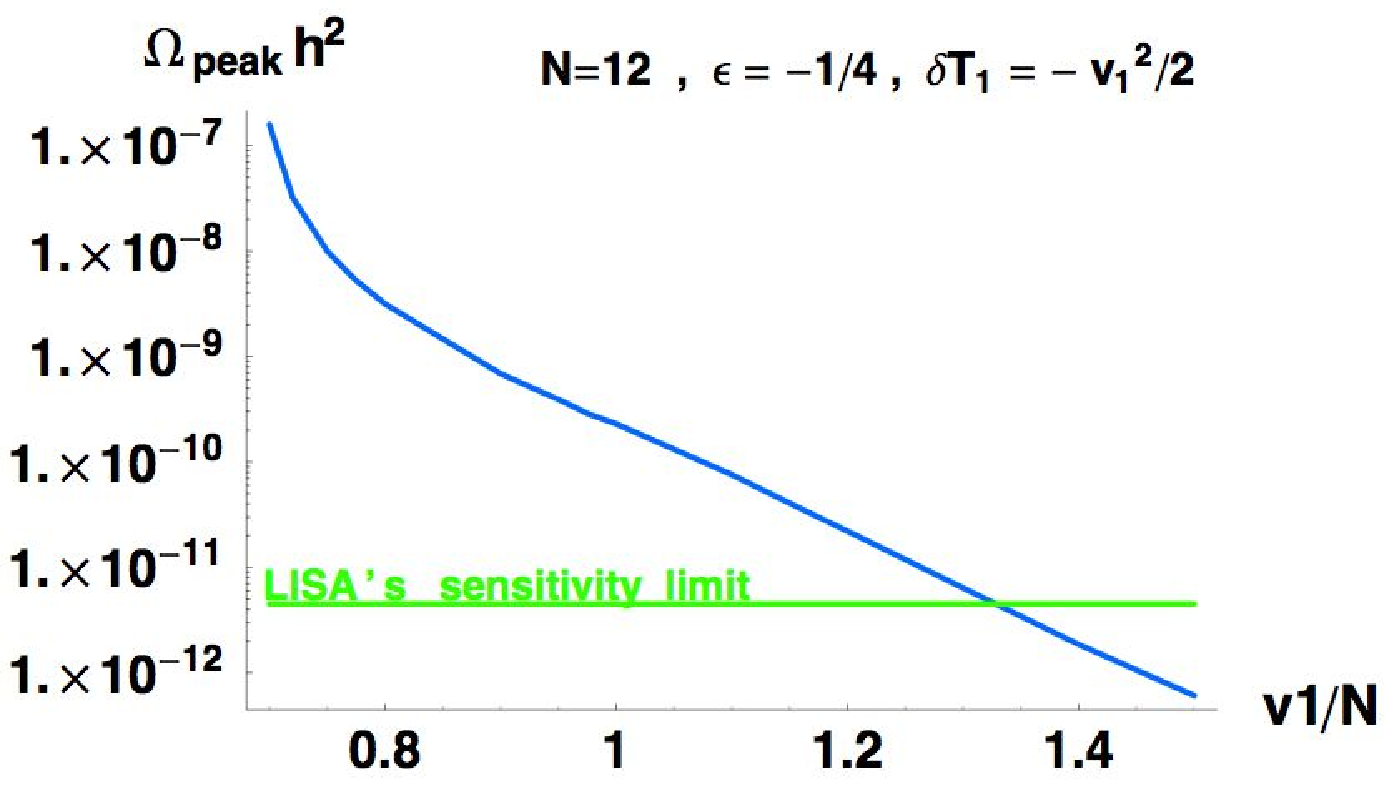}
\includegraphics[height=4.75cm,width=8.1cm]{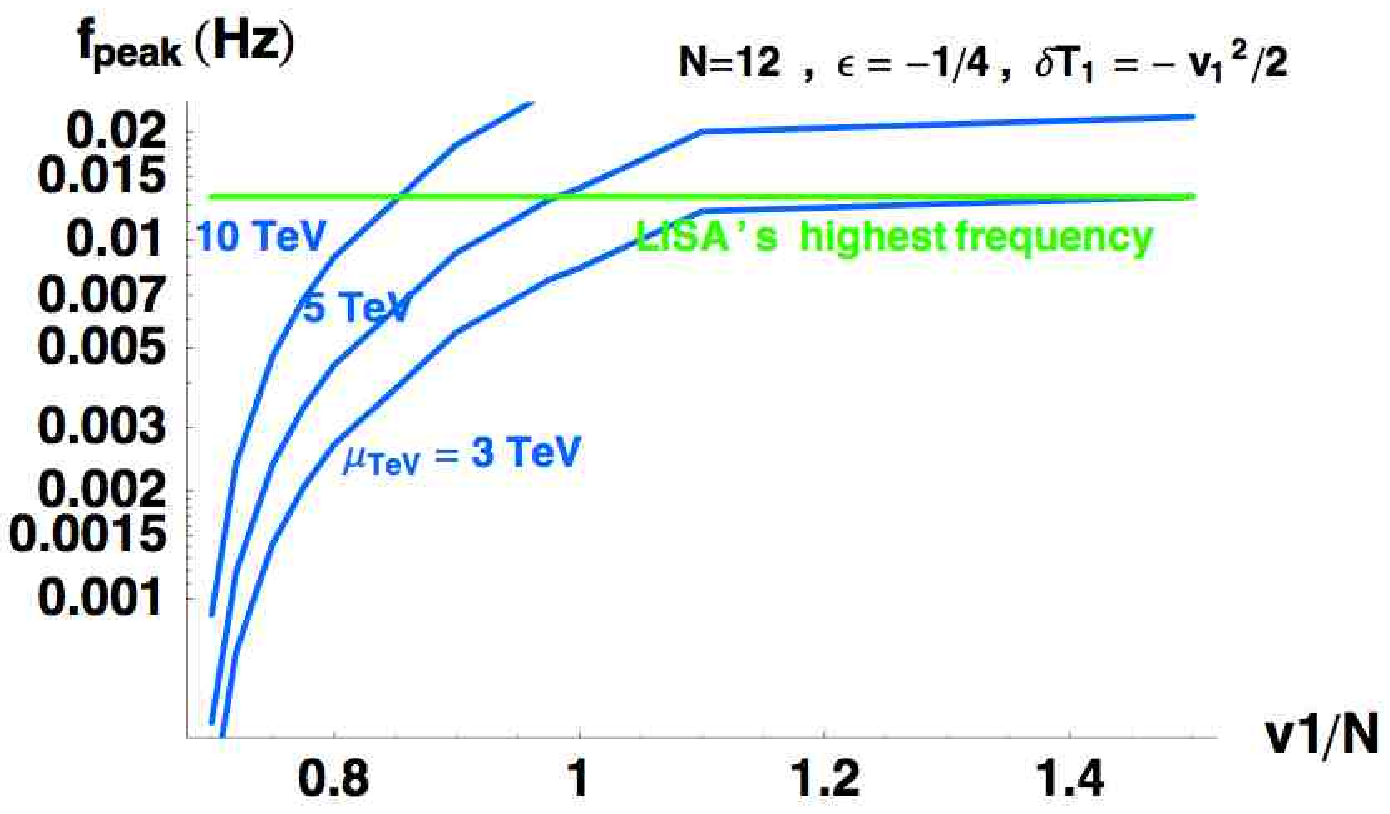}
\caption{  $\alpha$ and $\beta/H$ determine the spectrum of gravitational waves. They are calculated
for some benchmark point: $\epsilon=-0.25$, $N=12$, $\delta T_1=-0.5 v_1^2$.
The smallest values of $v_1/N$ correspond to a large amount of supercooling i.e. a small value of the ratio $T_n/T_c$.
This ratio varies between 0.23 for  $v_1/N=0.7$ to 0.87 for $v_1/N=1.5$.
Assuming that $\mu_{\mbox{\tiny TeV}}=5$ TeV, the corresponding nucleation temperatures are in the range 490 GeV -- 2700 GeV. We also show
 the peak frequency $f_{\mbox{\tiny peak}}$ of the gravitational wave signal from turbulence  and $\Omega_{\mbox{\tiny peak}} h^2$. The peak frequency depends on
$\mu_{\mbox{\tiny TeV}}$, while $\alpha$, $\beta/H$  and
$\Omega_{\mbox{\tiny peak}} h^2$ do not. As shown in Fig.~\ref{negativespectrum}, this can lead to a spectacular signal at LISA and/or BBO.}
 \label{alphabeta}
\end{center}
\end{figure}

 Interestingly, $\alpha$ and $\beta/H$ do not depend explicitly on the
 parameters of the GW potential, $\epsilon$,
 $v_1$, and $\delta T_1$ but only on  $a=T_n/T_c$. Of course, $T_n$ and
 $T_c$ implicitly depend on
 $\epsilon$, $v_1$, and $\delta T_1$.
Again, we have not included the extra contribution from the $T$-dependence in the RS potential.
This would lower the value of the latent heat but also lower the value of
$T_c$ and we expect that the effect on
 $T_n/T_c$ and thus $\alpha$ should not be significant. It could change the calculation of $\beta/H$ however.

From Figure \ref{alphaversusN}, it is clear that $\alpha$ can
be larger than 1 in much of the parameter regime. In particular, it gets very large if
$T_n/T_c<1/3^{1/4}$, when $O(4)$ symmetric bubbles have to  be considered.
 This justifies the approximation that bubble expansion proceeds via
detonation.
 We expect  large signals at LISA, whose sensitivity requires at least
$\alpha\gsim 0.2$ and
$\beta/H \lsim 1000$ for observability of the
 signal \cite{Nicolis:2003tg,Grojean:2006bp} (for a general detectability analysis at LISA and BBO in the $(\alpha, \beta/H)$ plane, for any given temperature, see \cite{Grojean:2006bp}).

 All predictions are functions of the nucleation
temperature $T_n$ which can be
 computed in the $\epsilon, v_1$ plane.
We will  present  in Fig.~\ref{positiveepsilon} the region of this plane in which the phase transition can take place.
 However, there will be strong constraints from perturbativity and
back-reaction which we discuss in the following section, and that will reduce the region
 of parameter space where predictions of large signals are reliable. We address these constraints
in the next section.

\begin{figure}[!htb]
\begin{center}
\includegraphics[height=6.cm,width=8.1cm]{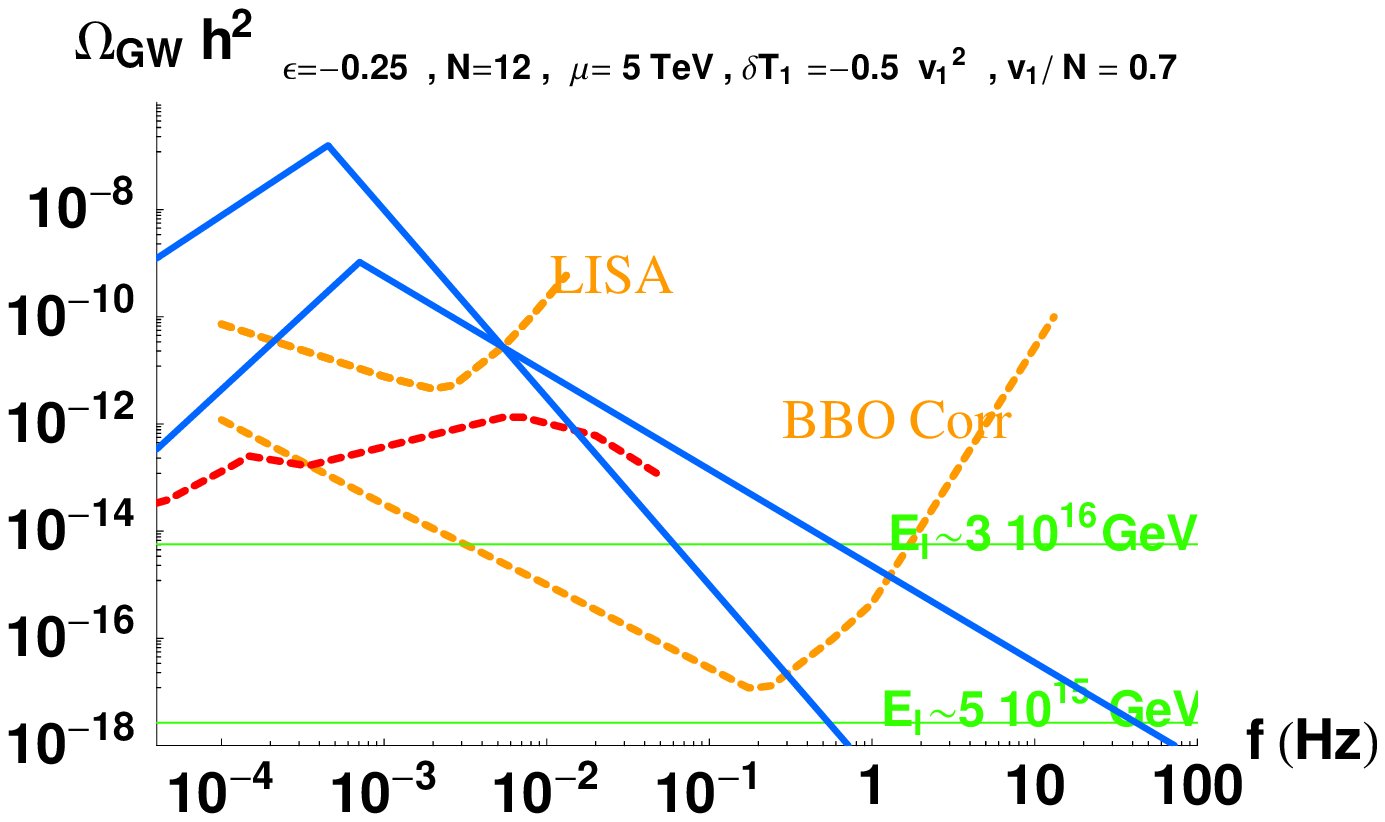}
\includegraphics[height=6.cm,width=8.1cm]{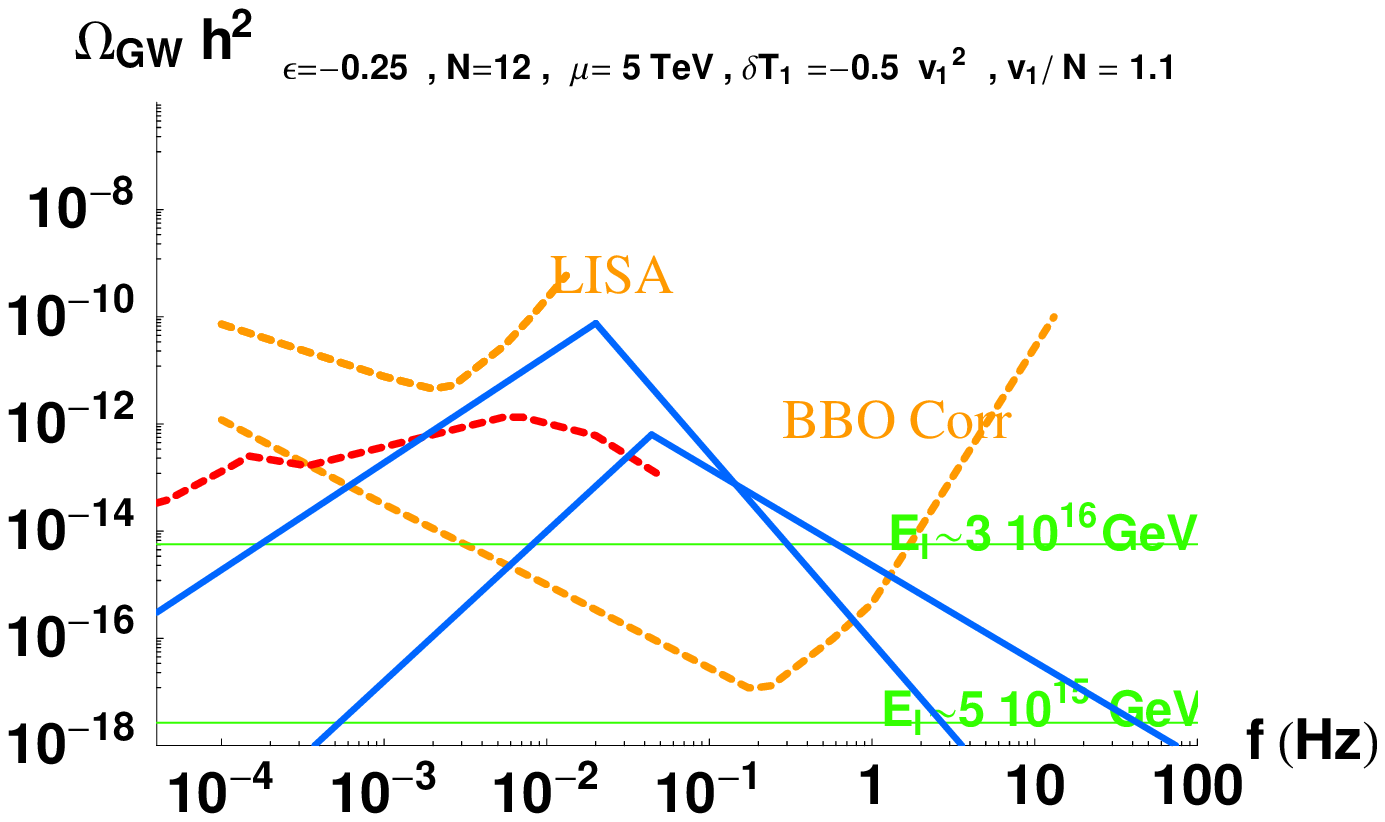}
\caption{Relic energy density in gravitational waves produced at the phase transition, $\Omega_{GW} h^2$, versus frequency for two values of $v_1$ corresponding to the benchmark point of Fig.~\ref{alphabeta}. The upper dashed orange line is the LISA sensitivity. The lower one is the sensitivity for the second generation Big Bang Observatory.  For each set of spectra, the first peak is from turbulence while the second peak is from bubble collisions. $v_1=0.7 N$ leads to a very strong phase transition ($\alpha=26$, $\beta/H=21$) at a temperature $T_n= 490$ GeV and a huge signal at LISA. For larger $v_1$, $\beta/H$ is higher and the peak frequencies are shifted to larger values. At some point, the signal is no more visible by LISA but still observable at BBO. The $v_1=1.1 N$ spectrum corresponds to $\alpha=1.6$, $\beta/H=350$ and a nucleation temperature $T_n=1830$ GeV.  The horizontal lines are gravitational wave spectra expected from inflation, for comparison (BBO is mainly planned to detect these gravitational waves), for two different scales of inflation. The dashed red curve is the expected irreducible background due to white dwarf binaries \cite{Farmer:2003pa}. }
\label{negativespectrum}
\end{center}
\end{figure}

We now show some typical values of the quantities that are relevant to the gravitational wave
signal as a function of $v_1$.
Fig.~\ref{alphabeta} shows  the values of $T_n/T_c$, $T_n/\mu_{\mbox{\tiny TeV}}$, $\alpha$, $\beta/H$ as well as the characteristic quantities of the gravitational wave signal obtained for a benchmark point $\epsilon=-0.25$, $|\delta T_1|=v_1^2/2$. The predicted gravitational wave spectrum for two particular values of $v_1$ are presented in Fig.~\ref{negativespectrum}.
In this region of parameter space
there is a huge signal of gravitational waves.
While relatively large values of $v_1/N$ are needed for the phase transition to take
place, once we are in the region where
 there is a phase transition, the smallest allowed values of $v_1/N$ lead to
the largest signals
(as much as three orders of magnitude above LISA's sensitivity) as they correspond
to the largest amount of supercooling.
The largest values of $v_1$ lead to larger characteristic frequencies no more
observable at LISA but within the
sensitivity range of BBO. These larger values of $v_1$ are worse for perturbativity in any case.  Note also that in this case, the temperature of the transition is not far from  the KK scale given by $\mu_{\mbox{\tiny TeV}}$.
For the particular region of parameters of Fig.~\ref{alphabeta}, nucleation takes place via 0(4) bubbles for $v_1/N<1.1$ and via 0(3) bubbles for larger $v_1/N$.
 Since $S_3/T$ and $S_4$ scale as a positive power of $N$, large $N$ suppresses nucleation and we present results for the maximum value of $N$, $N=12$, beyond which the phase transition cannot complete.  Relatively low values of $N$ are reasonable to explore since if $N$ were much larger, one risks losing asymptotic freedom. In any case, the size of the gravitational wave signal does not depend explicitly on $N$ but only on the amount of supercooling, which means that we prefer to live close to the limits of the region in the ($\epsilon$, $v_1/N$) parameter space where the transition completes. While $N$ is very important to determine the region of parameter space where the phase transition takes place, once $N$ is fixed, the size of the GW signal depends only on how far we are from the limits of the underlying region, which depend solely on $v_1/N$ and $\epsilon$. And different values of $N$  can lead to the same size for the signal. Therefore, the choice $N=12$ in Fig.~\ref{alphabeta} and Fig.~\ref{negativespectrum} does not really matter.

Note that the results depend on the scale $\mu_{\mbox{\tiny TeV}}$ only through the frequency of the signal.
Indeed, $\alpha$ does not depend on $\mu_{\mbox{\tiny TeV}}$.  $S_3/T$, $S_4$ and $\beta/H$ do not depend on $\mu_{\mbox{\tiny TeV}}$ explicitly.
 It is only when they are evaluated at the nucleation temperature that some weak logarithmic dependence on the energy scale of the phase transition appears.
 Therefore, the amplitude of the signal $\Omega h^2$ virtually does not depend on
 $\mu_{\mbox{\tiny TeV}}$.
On the other hand, the temperature of the transition and thus the peak frequency of the signal  is proportional to $\mu_{\mbox{\tiny TeV}}$. In the particular example of Fig.~\ref{alphabeta}, we plot the peak frequency in the three cases  $\mu_{\mbox{\tiny TeV}}=3, 5, 10$ TeV.  This shows that it is possible to see the peak of the turbulence signal at LISA for $\mu_{\mbox{\tiny TeV}}$ as large as $\sim$10 TeV. This does not mean that higher values of $\mu_{\mbox{\tiny TeV}}$ are not accessible. In fact, for some  $\epsilon, v_1$ values, it could still be possible to see at LISA the low frequency tail of the signal for $\mu_{\mbox{\tiny TeV}} $ as large as a few hundreds of TeV. This can easily be seen by translating the blue peak of the first plot of Fig.~\ref{negativespectrum} to the right by a factor  $\mu_{\mbox{\tiny TeV}}/5$ TeV.

\section{Perturbativity Constraints}
\label{sec:perturbativity}

We have seen in the last section that the first order phase transition from
AdS-Schwarschild to RS1 geometry could lead to spectacular gravitational
wave signals if the first order phase transition completes at about the TeV scale.
In this section, we consider the constraints from
perturbativity and the limits on our perturbative analysis. We express our answers in terms of $N$ of a CFT but that is directly related to the ratio of mass scales through our assumed relation $(ML)^3=N^2/16 \pi^2$. 
 In order to trust the background AdS solution, we impose  $1/(16 \pi^2(ML)^3)\lsim 1$, which leads to the weak bound  $N \gsim 1$.  Most phenomenological models have relatively small values of $N$. As we will see, for $N\gsim 12$, the transition cannot complete.
Small $v_0$ and small $\epsilon$, which are needed for perturbativity,  suppress the tunneling rate. We now consider the constraints on these parameters.
   
   Essentially there is a trade-off between good ranges of each of the parameters. Smaller $N$ leads to better values of $v_0$ and $v_1$.
 The constraints on $v_0$ and $v_1$ come from avoiding too large a back-reaction to the AdS energy of the five-dimensional theory, as well as imposing the radion-dominance assumption that was critical to the four-dimensional analysis.   When a vev is too big, there can be large corrections to the GW potential from KK modes and large corrections to the radion kinetic term due to mixing with the GW scalar.
According to \cite{Kofman:2004tk}, the last two sources can be acceptable however. That is, when the full mass matrix and kinetic matrix are diagonalized, one can still find dominance of the light radion. Of course, in that case, we are not guaranteed that the radion potential takes the form we assumed and we are not guaranteed the transition remains first order. However, we see no reason to assume that in all cases the first order phase transition would disappear, but of course we don't know for sure without a more complete analysis in the large vev regime.

Here we will focus on the constraint on small back-reaction to the energy. Notice that even this is potentially stronger than is necessary.
Imposing the constraint $v_0, v_1<N/(4\pi)$ i.e. requiring that the vevs of the Goldberger-Wise field are smaller than $M$, the 5d Planck scale,
clearly suppresses the back-reaction. As \cite{DeWolfe:1999cp} have shown, small back-reaction to the energy is not necessarily essential to maintaining the hierarchy\footnote {Note that their model with large back-reaction would also have a large vev that could induce corrections to the potential that were not discussed.}.
 Nonetheless, it would be best to have small back-reaction so that the original AdS space analysis can be trusted. We view these plots as indicative of the proximity of the perturbativity limits.

The stress tensor for the Goldberger-Wise field $\phi$ is
\be
T^{\phi}_{MN}=-g_{MN}[-(\partial \phi)^2/2-m^2 \phi^2/2] -\partial_M \phi\partial_N \phi
\ee
Using the metric  $ds^2=(dx_{\mu} dx^{\mu}+dz^2)/(k^2z^2)$ and
$\phi(z)=A z^{4+\epsilon}+Bz^{-\epsilon}$
where
\be
A=z_1^{-4-\epsilon}k^{3/2}\frac{v_1-v_0(z_0/z_1)^{\epsilon}}{1-(z_0/z_1)^{4+2\epsilon}} \ \ \ \ \ , \ \ \
B=z_0^{\epsilon}k^{3/2}\frac{v_0-v_1(z_0/z_1)^{4+\epsilon}}{1-(z_0/z_1)^{4+2\epsilon}}
\ee
 We find
\bea
T^{\phi}_{\mu\nu}(z)=2\eta_{\mu\nu}[A^2 z^{6+2\epsilon}(4+3\epsilon) + \epsilon B^2z^{-2-2\epsilon}]
\eea
When we  compare this with the energy momentum tensor due to the bulk cosmological constant $\Lambda_5=-24 M^3 k^2$
\be
T^{\Lambda}_{\mu\nu}=(kz)^{-2}\eta_{\mu\nu}\Lambda_5
\ee
we find the condition
\be
v_0^2< 12 (ML)^3/|\epsilon|  \   \   \  i.e.  \   \   \ \frac{v_1}{N}<\sqrt{\frac{3}{4}}
\frac{1}{\pi \sqrt{|\epsilon|}X}\left(\frac{\mu_{\mbox  {\tiny TeV}}}{\mu_0}\right)^{\epsilon}
\label{UVBR}
\ee
on the Planck brane and
\be
v_1^2< \frac{12 (ML)^3}{|(4 + 3 \epsilon)(1- X)^2+ \epsilon X^2|} \   \ \ i.e.  \   \   \
\frac{v_1}{N}<\sqrt{\frac{3}{4}}\frac{1}{\pi}\frac{1}{\sqrt{|(4 + 3 \epsilon)(1- X)^2+ \epsilon X^2|}}
\label{IRBR}
\ee
on the TeV brane,
where $X\equiv1+\frac{\epsilon}{4}\pm \frac{1}{2}\sqrt{-\frac{\delta T_1}{v_1^2}+\epsilon+\frac{\epsilon^2}{4}} $, $X \in [0,2]$ appears in the relation between $v_0$ and $v_1$, Eq.~(\ref{mutev}).

If instead of evaluating the constraint on the TeV and Planck branes we integrate it over $z$, we get
the same as eq. (\ref{UVBR}).
For $|\delta T_1|=0$, $X=1$ and the conditions (\ref{UVBR}) and (\ref{IRBR}) respectively  become
$v_1^2< 12 (ML)^3/|\epsilon|$ and
$v_0^2< 12 (ML)^3/|\epsilon|$
 but for $|\delta T_1|=4 v_1^2$, the back reaction on the IR brane leads to  the stronger constraint
\be
v_1^2< 3 (ML)^3
\ee
Notice these constraints are readily understood when $\delta T_1$ is zero. If $v_1^2$ or $v_0^2$ (normalized to a dimensionless quantity through factors of $k$) are too large compared to the Planck scale, the expansion in powers of the Goldberger-Wise field would be invalid.  Furthermore, if the Goldberger-Wise field enters only through the kinetic and mass terms, we have the approximate condition $m^2\phi^2/2  \sim 2 k^2 \epsilon \phi^2 <24 k^2 M^3$, we get the conditions above for zero $\delta T_1$.  With the $\delta T_1$ turned on and large, it already implies some back-reaction near the TeV brane.
\begin{figure}[!htb]
\begin{center}
\includegraphics[height=7.2cm,width=8.1cm]{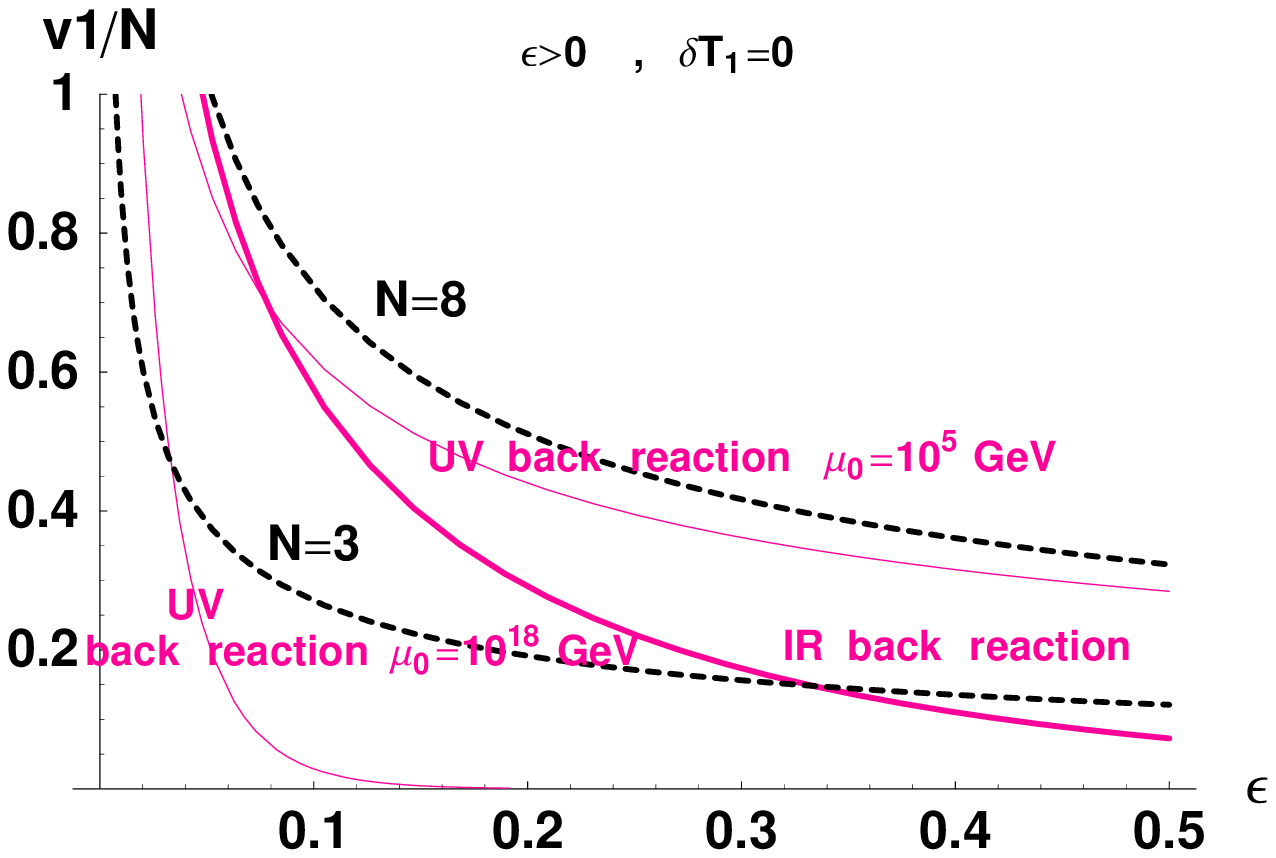}
\includegraphics[height=7.2cm,width=8.1cm]{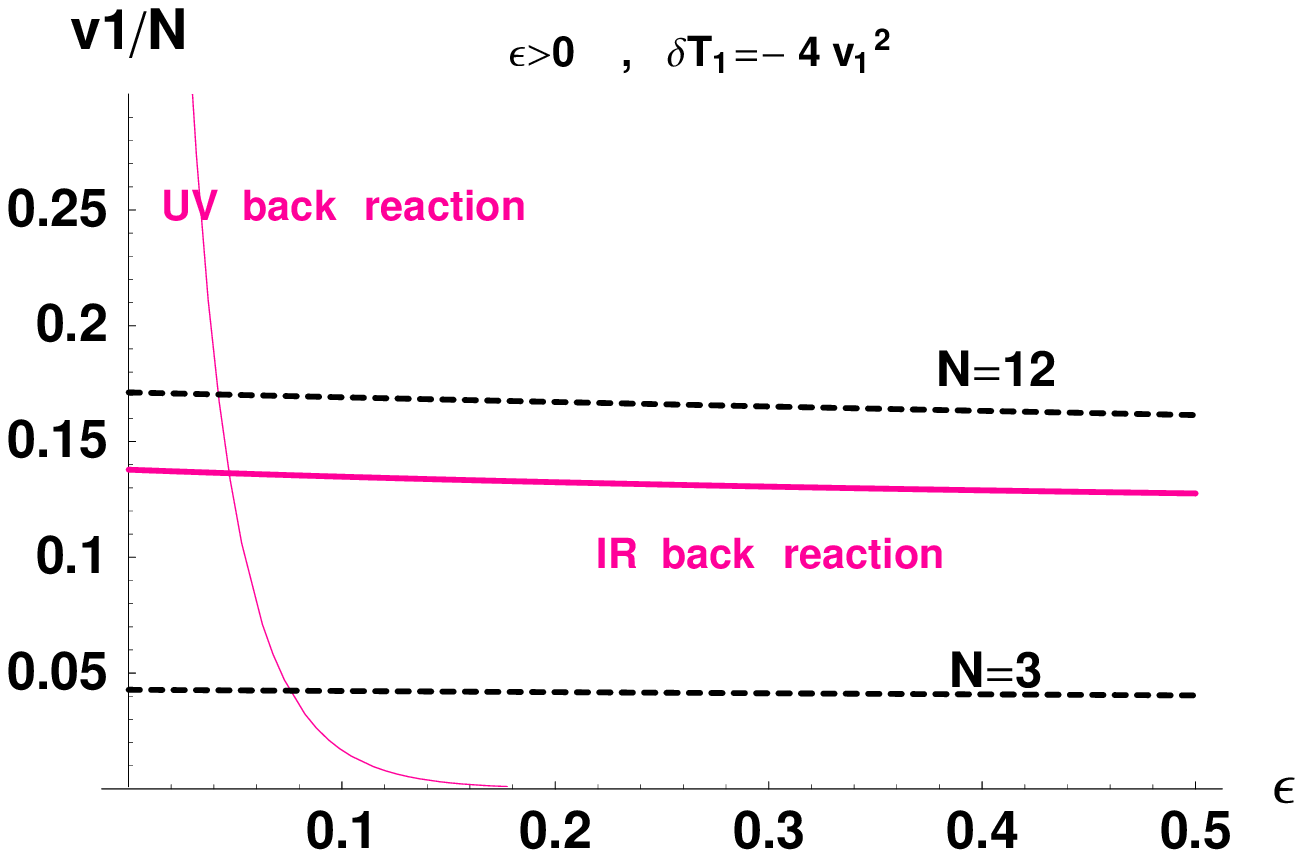}
\includegraphics[height=7.2cm,width=8.1cm]{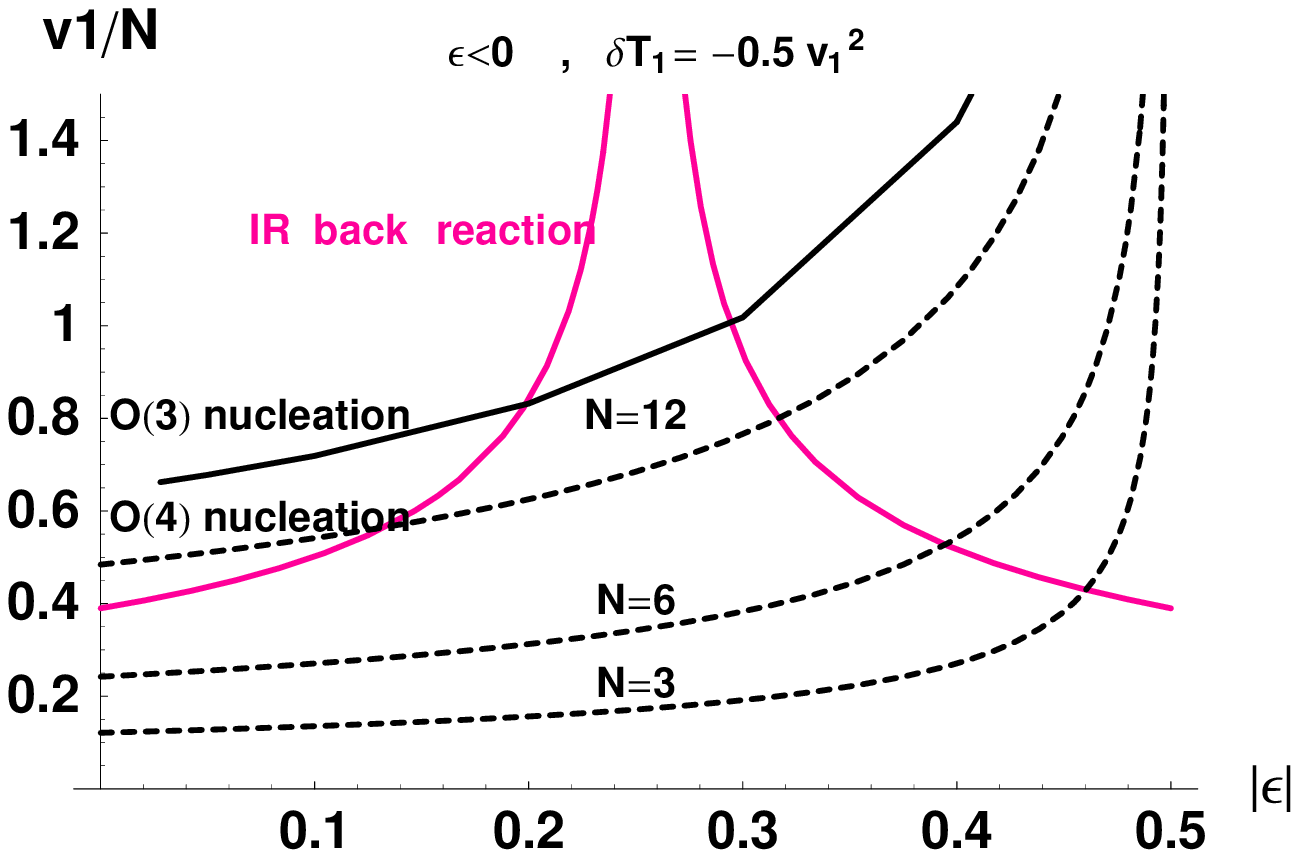}
\includegraphics[height=7.2cm,width=8.1cm]{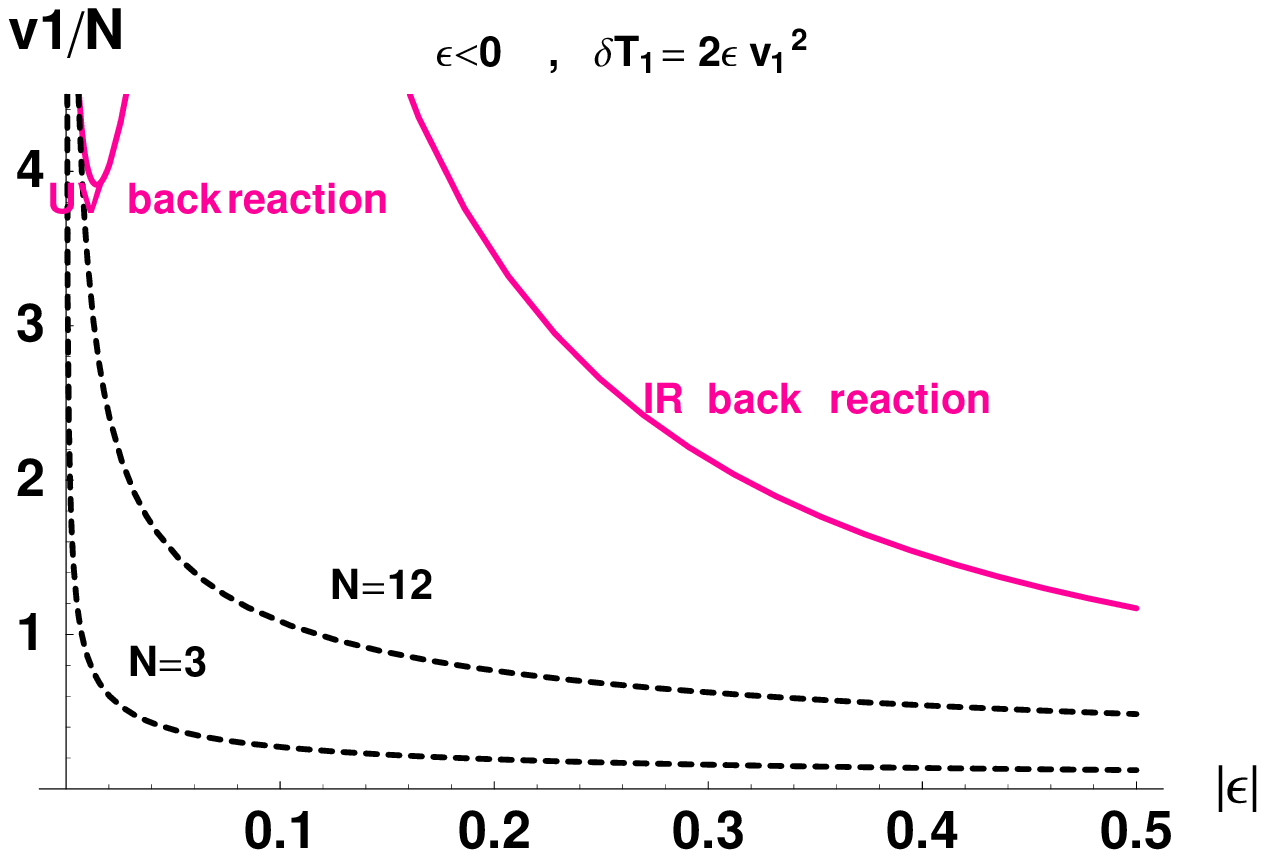}
\caption{$\epsilon >0$ (top) and  $\epsilon <0$ (bottom). The dashed black line delimits the region where  nucleation of $O(4)$ symmetric thick-walled bubbles is possible.  We have to be above this line for $O(4)$ nucleation to take place. The nucleation contour for thick-walled $O(3)$-symmetric bubbles is typically slightly above (see the black solid line on the third plot).  Close to the nucleation line, the phase transition is very strong. The pink lines delimit the region where back-reaction is important.  Regions that satisfy the back-reaction constraint are below these lines. Almost no region remains for $\epsilon >0$,  but with $\epsilon <0$, some regions satisfy the constraints. We have shown how decreasing $N$ enlarges the parameter space.}
 \label{positiveepsilon}
\end{center}
\end{figure}

We plot these constraints in Fig.~\ref{positiveepsilon}.
For positive $\epsilon$, the IR back reaction constraint is too big and nucleation never takes place in a regime where we can ignore the back reaction.
 Because positive $\epsilon$ requires
$v_0>v_1$, the UV backreaction condition
  is usually stronger.  However, from a strictly phenomenological
perspective, this constraint
 is not essential as it comes from the requirement that the RS model
applies all the way to the Planck scale.
 We can relax the constraint  with a lower cutoff  and assume some
unknown ultraviolet physics
 beyond that scale. In Figure \ref{positiveepsilon} , we show how the
viable region of parameter space in
 the $\epsilon,v_1$ plane increases as we lower the UV scale. However, even a UV scale at $10^5$ GeV is not enough to make nucleation in the perturbative region.

\subsection{Negative $\epsilon$}

We find that results are more favorable for $\epsilon <0$. This is not because nucleation is easier
 but because the back reaction constraints are  weaker.
The point is that the constraints on $v_0$ and $v_1$ are parametrically comparable. But for $\epsilon <0$,  we have $v_0<v_1$ (see Eq. \ref{mutev}), so the perturbativity constraint on $v_0$ is generally satisfied once the constraint on $v_1$ is met\footnote{A bound on $v_0$ comparable to (\ref{UVBR}) is obtained in the particular example of Ref.~\cite{Kofman:2004tk},  where the back-reaction parameter is defined as $l= 4 \pi  v_0/N$.
To guarantee that the radion and KK modes are heavy enough and not too strongly coupled to SM fields on the TeV brane, they need $l \lsim 10$ $i.e.$ $v_0/N \lsim 1$. Since $v_1\sim v_0 (\mu_{\mbox{\tiny TeV}}/\mu_0)^{\epsilon}$, this constraint is extremely strong for $\epsilon >0$ and incompatible with our nucleation condition but for $\epsilon <0$, it is readily satisfied.}.
As for the IR constraint (\ref{IRBR}), we find regions of parameter space where it is relaxed, for instance if $\delta T_1= 2 \epsilon v_1^2$ where the right hand side of the inequality  (\ref{IRBR})  blows up to infinity. This follows from the right hand side of eq (\ref{UVBR}) which blows up at large $|\epsilon|$ as $(\mu_0/\mu_{\mbox{\tiny TeV}})^{|\epsilon|}$.
 All this is illustrated in Fig.~\ref{positiveepsilon}. Of course we are ignoring higher order terms so we take this result as simply indicative of the fact that although nucleation takes place at or above the perturbativity limit in general for $N>12$,  the leading term and higher order terms might conspire to be small.
 
 Notice that the position of the (pink) lines delimiting the perturbativity regions are $N$-independent while the nucleation (black) lines get shifted to lower values of $v_1/N$ at smaller $N$. For instance, for
$N\sim 3, 4$, it is possible to obtain nucleation at $v_1/N \sim 0.1$ rather than at $v_1/N \sim 1$ and therefore to enlarge the region of parameter space where the phase transition completes in the regime of small back reaction, as illustrated on Fig.~\ref{positiveepsilon}.

 There is, however, an additional constraint for the negative $\epsilon$ case as the operator that breaks conformal invariance gets strong in the IR. Below the scale of strong coupling $\Lambda$, the conformal picture is spoiled so perturbativity also requires that
the nucleation temperature exceeds  $\Lambda$, where $\Lambda$ should correspond to the value of $\mu$ at which  $v_1$ takes its maximum value consistent with perturbativity. We find this constraint essentially coincides with $v_1/N\sim1$.   Since the strong coupling scale
is not a precisely defined quantity, we would hope that strong coupling effects are not big before nucleation. The first order phase transition might be possible, even with strong coupling effects included. However, if strong coupling effects are important, we can no longer be confident about the order of the phase transition.

\section{Conclusion}

There can be a significant signal for gravitational waves from the phase transition from AdS-Schwarschild to RS1.
Predictions vary by orders of magnitude depending on the region of parameter space associated with the scalar potential stabilizing the radion but if the transition completes, the signal is likely to be significant.  The KK scale that can be probed is much higher than at colliders. An IR scale as large as a few tens of TeV (even hundred) is in principle reachable at LISA, depending on the region of parameter space. The uncertainties in the computation come from the unknown temperature-dependence of the potential and the unknown back-reaction effects.

We have focused on the gravitational wave signal that is indicative of any first order phase transition; we have not included any features peculiar to this theory. That means that even if this signal is measured, we cannot definitively state that it was from the phase transition we considered. However, there have been extensive investigations of existing weak-scale models, primarily with the aim of studying electroweak baryogenesis, and no model has yet been found with such a strong transition ($i.e.$ with $\alpha \gg 1$).
For example, the phase transition we have considered can be much stronger than the usual EW phase transition, which is usually constrained by the experimental bound on the Higgs mass. In our case, there are no such bounds. The radion sector is much less constrained and
this offers the possibility of a strong phase transition.
Although it would not be conclusive, a large LISA signal of the sort we have discussed is likely to be associated with an RS1 cosmological phase transition.

This gravitational wave signature has the advantage of being common to all RS1 models. This  is to be contrasted with collider signatures which depend on the details of the model such as the localization of the Standard Model fermions and gauge fields in the bulk of $AdS_5$. On the other hand, the signal strongly depends on the radion sector which to some extent can be probed  by collider experiments.
We are finding that the phase transition can proceed only in the regime of large back reaction. According to Ref.~\cite{Kofman:2004tk}, the radion couplings are expected to be large in this case. This could indeed imply the possibility of testing the strength of the transition at colliders.

We emphasize that the uncertainties in the calculation are primarily due to the proximity  to  the perturbativity limits of the calculation. However, if the phase transition completes, it is clear that the signal is likely to  be large.
 It will be interesting to see how far these results can be extended
into the nonperturbative regime. It will also be interesting to investigate other possibilities for weak scale physics and gravitational waves.

\section*{Acknowledgements}
We thank Riccardo Rattazzi, Jose Ramon Espinosa, Alberto Nicolis, Jared Kaplan,
Natalia Toro,
Philip Schuster, C. Caprini for discussions.
Lisa Randall acknowledges the NSF support  through grants  PHY-0201124 and PHY-0556111.

\end{document}